\documentclass[%
 reprint,
 amsmath,amssymb,
 aps,pre
floatfix,
]{revtex4-1}

\usepackage{graphicx}
\usepackage{dcolumn}
\usepackage{bm}
\usepackage{mathtools}
\usepackage{color}
\usepackage{xspace}

\newcommand{\DISS}{\omega}
\newcommand{\TOT}{\omega_{\text{tot}}}

\newcommand{\RCF}{k_{ij}^+}
\newcommand{\RCR}{k_{ij}^-}
\newcommand{\RCFONETWO}{k_{12}^+}
\newcommand{\RCFTWOONE}{k_{21}^+}
\newcommand{\RCFTWOTHREE}{k_{23}^+}
\newcommand{\RCFTHREEONE}{k_{31}^+}
\newcommand{\RCRONETWO}{k_{12}^-}
\newcommand{\RCRTWOONE}{k_{21}^-}
\newcommand{\RCRTWOTHREE}{k_{23}^-}
\newcommand{\RCRTHREEONE}{k_{31}^-}

\newcommand{\DOPT}{\omega^*}
\newcommand{\DOPTPS}{\omega^{*,\text{FL}}}
\newcommand{\DOPTBR}{\omega^{*,\text{RL}}}

\newcommand{\DELD}{\Delta\omega}
\newcommand{\BARE}{k_{ij}^0}

\newcommand{\BAREONETWO}{k_{12}^0}
\newcommand{\BARETWOONE}{k_{21}^0}
\newcommand{\BARETWOTHREE}{k_{23}^0}
\newcommand{\BARETHREEONE}{k_{31}^0}
\newcommand{\MD}{\text{d}}
\newcommand{\powerstrokenew}{forward labile\xspace}
\newcommand{\brownianratchetnew}{reverse labile\xspace}
\newcommand{\Powerstrokenew}{Forward labile\xspace}

\newcommand{\PowerStrokenew}{Forward Labile\xspace}
\newcommand{\BrownianRatchetnew}{Reverse Labile\xspace}
\newcommand{\powerstrokenewacr}{FL\xspace}
\newcommand{\brownianratchetnewacr}{RL\xspace}

\begin{document}

\preprint{APS/123-QED}

\title{Allocating dissipation across a molecular machine cycle to maximize flux}

\author{Aidan I. Brown}
\email{aidanb@sfu.ca}
\author{David A. Sivak}
\email{dsivak@sfu.ca}
\affiliation{Department of Physics, Simon Fraser University, Burnaby, BC, V5A1S6 Canada}
\date{\today}

\begin{abstract}
Biomolecular machines consume free energy to break symmetry and make directed progress. Nonequilibrium ATP concentrations are the typical free energy source, with one cycle of a molecular machine consuming a certain number of ATP, providing a fixed free energy budget. Since evolution is expected to favor rapid-turnover machines that operate efficiently, we investigate how this free energy budget can be allocated to maximize flux. Unconstrained optimization eliminates intermediate metastable states, indicating that flux is enhanced in molecular machines with fewer states. When maintaining a set number of states, we show that---in contrast to previous findings---the flux-maximizing allocation of dissipation is not even. This result is consistent with the coexistence of both `irreversible' and reversible transitions in molecular machine models that successfully describe experimental data, which suggests that in evolved machines different transitions differ significantly in their dissipation.
\end{abstract}

\maketitle

\section{Introduction}
Biomolecular machines, typically composed of protein complexes, perform many roles inside cells, including cargo transport and energy conversion~\cite{kolomeisky13}. These microscopic machines operate stochastically~\cite{astumian94}, but must on average make forward progress to fulfill their cellular roles, a functional requirement that according to the Second Law imposes a free energy cost~\cite{machta15}.

Biomolecular machines typically make use of the free energy stored in nonequilibrium chemical concentrations, which are in turn maintained by other cellular machinery
~\cite{boyer97}.
The free energy consumed over a forward machine cycle equals the free energy difference between the chemical reactants and products~\cite{philips12}, which sets the maximum available dissipation `budget' for a cycle.

Theoretical studies have found that under a variety of criteria, an even allocation of dissipation across all transitions in a machine cycle is optimal~\cite{qian16,qian00,sauar96,johannessen05,oster00,hill81,yu07,anandakrishnan16,barato15,geertsema09}. However, many models parametrized to experimental  biomolecular machine dynamics contain effectively irreversible transitions~\cite{clancy11,visscher99,cappello07,xie06,abbondanzieri05,moffitt09,liu14}, suggesting that some transitions dissipate a large amount of free energy compared to the `reversible' transitions in the same cycle. 

The dissipation allocation generally affects the probability flux (also known as the current) through a molecular machine cycle~\cite{astumian15}. Flux reports on the machine output and thus is an important operating characteristic; indeed, the dependence of flux on alternative energy landscapes was recently proposed to explain the ubiquity of the rotary mechanism of ATP synthase~\cite{anandakrishnan16}.

We approximate molecular machine dynamics with stochastic transitions between discrete states
~\cite{kolomeisky13,keller00}
and examine how a fixed free energy dissipation budget should be allocated to a cycle's individual transitions to achieve maximal flux. We find that without further constraints, maximizing the flux effectively eliminates the free energy wells representing intermediate metastable states. When constrained to maintain a set number of intermediate metastable states, our central result is that flux is maximized when dissipation is unevenly allocated among the distinct cycle transitions.

Our result is consistent with the presence in the same cycle 
of both reversible and effectively irreversible transitions, and the 
substantially
different implied dissipations~\cite{clancy11,visscher99,cappello07,xie06,abbondanzieri05,moffitt09,liu14}. This suggests that understanding how forward progress is affected by a dissipation allocation may be useful for evaluating the design of biomolecular machines. Adjustment of the dissipation allocation of a biomolecular machine may be relatively easy to parsimoniously achieve, compared to broad-reaching changes such as to the fuel source or to the free energy of ATP hydrolysis.

\section{Models}
\label{sec:model}

\subsection{Discrete states}
\begin{figure}[tbp] 
	\centering
	\hspace{-0.0in}
	\begin{tabular}{cc}
		\hspace{-0.200in}\includegraphics[width=1.4in]{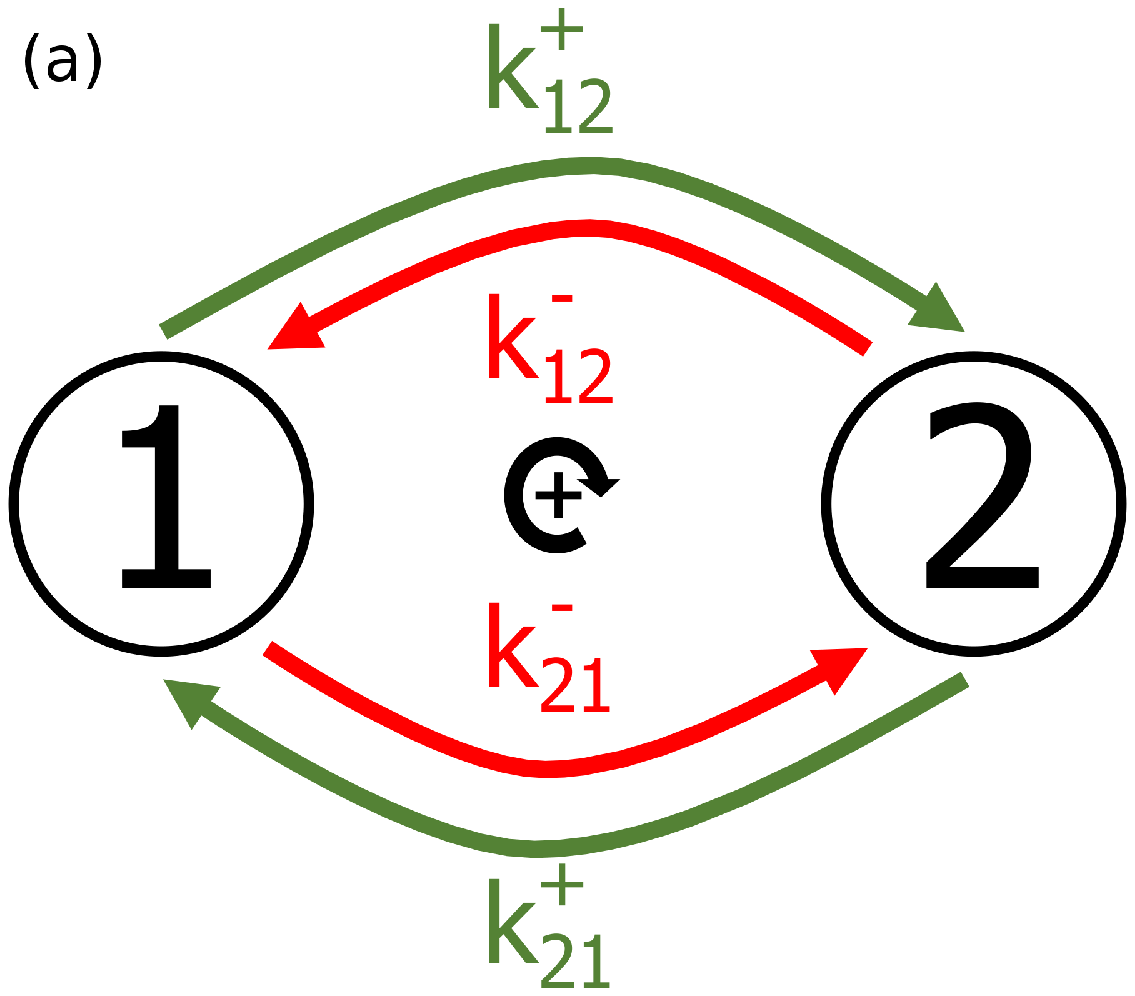} &
        \hspace{0.200in}\includegraphics[width=1.4in]{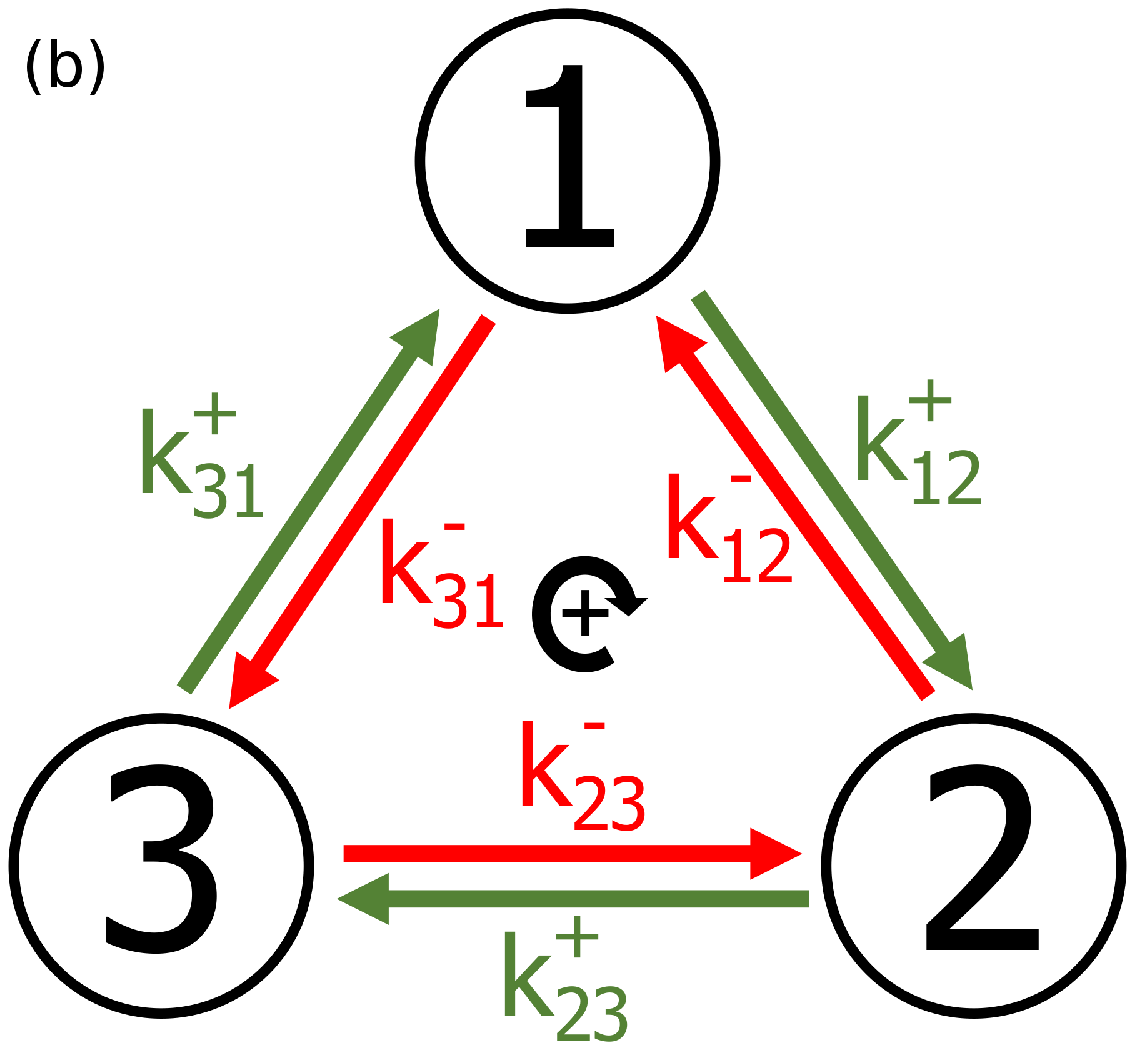}\\
	\end{tabular}
	\caption{\label{fig:Diagrams} 
{\bf Small discrete-state cycles.} (a) Two-state and (b) three-state cycles. Forward (clockwise, green) and reverse (counterclockwise, red) transitions occur along each pathway connecting two states, with respective rate constants $\RCF$ and $\RCR$.
}
\end{figure}

We consider two- and three-state cycles (Fig.~\ref{fig:Diagrams}), which have frequently been used to
model 
driven \emph{in vivo} systems,
such as myosin~\cite{kolomeisky03}, kinesin~\cite{clancy11}, phosphorylation-dephosphorylation cycles~\cite{qian06}, 
the canonical Michaelis-Menten scheme~\cite{philips12},
and various specific enzymes~\cite{qian06,hwang17}, 
For the two-state Michaelis-Menten scheme~\cite{philips12}, the first transition binds the substrate, while the second catalyzes the reaction of substrate to product, releases the product, and returns the enzyme to its original state. For a three-state kinesin model~\cite{clancy11}, the first transition binds ATP to the microtubule-bound head, the second steps the free head forward to bind the microtubule and release ADP, and the third hydrolyzes ATP and unbinds the newly rear head from the microtubule.

In our model cycle, for every forward rate constant $\RCF$ describing transitions from state $i$ to state $j$, there is a nonzero reverse rate constant $\RCR$ describing transitions from state $j$ to state $i$. Although some models of molecular machines describe certain transitions as `irreversible,' with a reverse rate of zero, this violates the principle of microscopic reversibility~\cite{astumian15, fisher99}.

A transition from state $i$ to state $j$ occurs at the forward rate $\RCF P_i$, with $P_i$ the probability in state $i$. Reverse transitions from state $j$ to state $i$ occur at rate $\RCR P_j$. (Note that the two-state cycle has both pathway 12 and pathway 21, representing distinguishable physical transition mechanisms, each with a forward and reverse direction.)

Typically, cellular cycles are driven by nonequilibrium concentrations of reacting chemical species, most prominently adenosine triphosphate (ATP), adenosine diphosphate (ADP), and inorganic phosphate ($\text{P}_{\text{i}}$)~\cite{philips12}. The free energy $\Delta G$ provided by ATP hydrolysis depends on the respective concentrations~\cite{philips12},
\begin{equation}
\Delta G = \Delta G_0 + k_{\text{B}}T\ln\frac{[\text{ADP}][\text{P}_{\text{i}}]}{[\text{ATP}]} \ ,
\end{equation}
where $\Delta G_0 \equiv -k_{\text{B}}T \ln\left([\text{ADP}]_{\text{eq}}[\text{P}_{\text{i}}]_{\text{eq}}/[\text{ATP}]_{\text{eq}}\right)$, $k_{\text{B}}$ is Boltzmann's constant, and $T$ is the temperature of the surrounding environment. Under physiological conditions, hydrolysis of ATP to ADP and $\text{P}_{\text{i}}$ provides free energy $|\Delta G| \sim 20 k_{\text{B}}T$~\cite{philips12}. From here on we set $k_{\text{B}}T = 1$ --- all free energies are in units of the thermal energy scale.

The ratio between the forward and reverse rate constants of a given transition path is fixed by its free energy dissipation $\DISS_{ij}$
~\cite{wagoner16,pietzonka16},
\begin{equation}
\label{eq:ratio}
\DISS_{ij} = \ln \frac{\RCF}{\RCR} \ . 
\end{equation}
(See Appendix.)
Without a bias $\DISS_{ij}$, the full forward and reverse rate constants $k_{ij}^{+/-}$ equal the `bare' rate constants $\BARE$.

\subsection{Free energy landscape}
\label{subsec:landscape}
\begin{figure}[tbp] 
	\centering
	\hspace{-0.0in}
	\begin{tabular}{c}
		\hspace{0.000in}\includegraphics[width=3.325in]{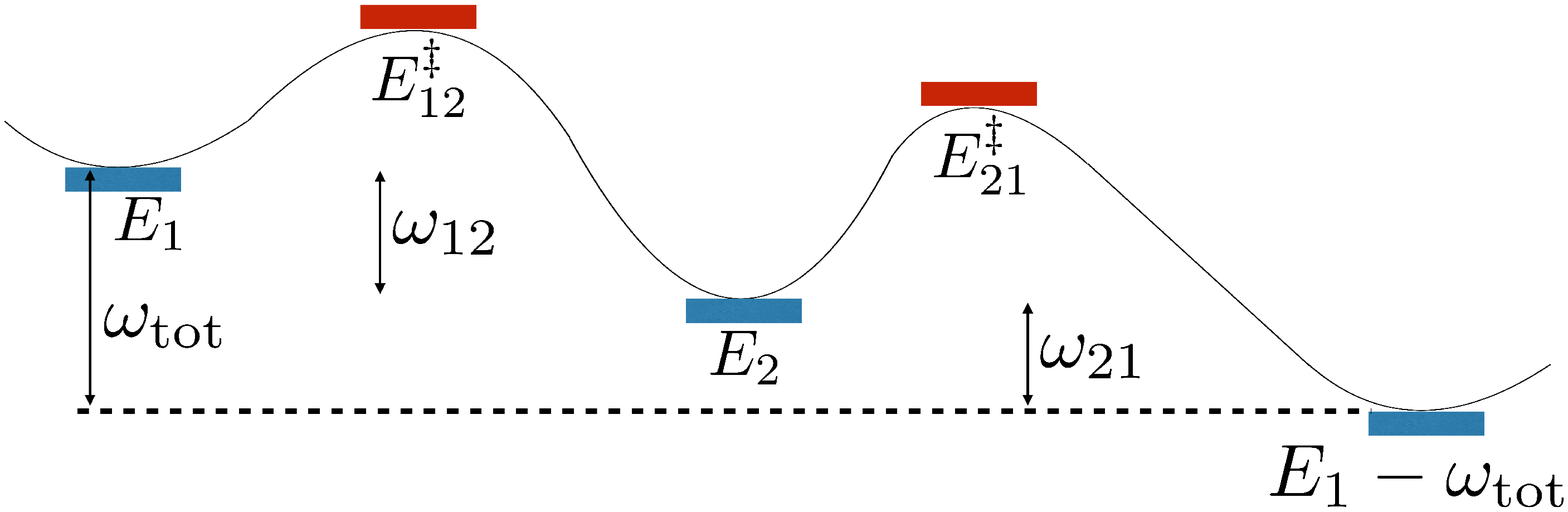}\\
	\end{tabular}
	\caption{\label{fig:Landscape} 
{\bf Free energy landscape} representing a two-state molecular machine. The left-most state (at free energy $E_1$) and the right-most state ($E_1 - \TOT$) represent the same stage of molecular machine operation, separated by one complete cycle. The middle state ($E_2$) represents an intermediate state, while $E^{\ddagger}_{12}$ and $E^{\ddagger}_{21}$ are the free energies of barriers between the states. $\DISS_{12}$ and $\DISS_{21}$ are the dissipations for each transition, which sum to $\TOT$, the dissipation budget for one cycle.}
\end{figure}

A discrete-state kinetic model of a machine cycle can also be equivalently represented by Arrhenius dynamics on a free energy landscape, with each free energy well representing a discrete state, and the free energy differences between barriers (at energies $E_{ij}^{\ddagger}$) and states (at energies $E_i$) determining the rate constants,
\begin{equation}
\RCF = \tau_{ij}^{-1}e^{-(E^{\ddagger}_{ij} - E_i)} \ \ \text{and} \ \  \RCR = \tau_{ij}^{-1}e^{-(E^{\ddagger}_{ij} - E_j)} \ ,
\end{equation}
with $\tau_{ij}$ a timescale accounting for effective diffusivity. 
The free energy budget $\TOT$ fixes the free energy difference between equivalent molecular machine states separated by one cycle. For example, Fig.~\ref{fig:Landscape} represents a two-state cycle, with dissipations $\DISS_{12} = E_1 - E_2$, and $\DISS_{21} = E_2 - (E_1 - \TOT)$.

The bare rates $\BARE$ are the rates in the absence of chemical driving. We restrict our attention to wells at equal free energy without chemical driving, so forward and reverse bare rates of a given transition are equal.
We allow bare rates to vary among the different transitions to account for differences in barrier heights, effective diffusivity, and all other dissipation-independent factors.

\section{Results}

We maximize the steady-state flux by allocating a fixed free energy budget $\TOT$ among the free energy differences $\DISS_{ij}$ between discrete states (\emph{i.e.}\ dissipation over discrete transitions), which determines the full rate constants $k_{ij}^{+/-}$ and hence the net steady-state flux $J$ (from here on, the flux), which for a two-state cycle is~\cite{hill77}
\begin{equation}
\label{eq:TwoStateFlux}
J = \frac{\RCFONETWO\RCFTWOONE - \RCRONETWO\RCRTWOONE}{\RCFONETWO + \RCRONETWO + \RCFTWOONE + \RCRTWOONE} \ .
\end{equation}

We first consider freely varying $E_2$, $E^{\ddagger}_{12}$, and $E^{\ddagger}_{21}$ (see Fig.~\ref{fig:Landscape}). When barriers are higher than states, decreasing the barrier energies always 
increases flux,
\begin{equation}
\label{eq:FreeBarrier}
\frac{\partial J}{\partial E^{\ddagger}_{12}} < 0 \ \ \text{and} \ \ \frac{\partial J}{\partial E^{\ddagger}_{21}} < 0 \ .
\end{equation}
(See Appendix for details.) Flux is maximized when the `barriers' are at or below the states, thus no longer acting as barriers. This is an intuitive result, that all else equal, faster transitions (due to lowered barriers) produce a higher flux~\cite{wagoner16}.

Increasing $E_2$ is equivalent to decreasing the dissipation $\DISS_{12}$ and increasing $\DISS_{21}$, and vice versa. For fixed
barrier energies $E^{\ddagger}_{12}$ and $E^{\ddagger}_{21}$, flux is maximized by increasing $E_2$ above the free energy of one of the barriers, effectively producing a one-state cycle (see Appendix).

Transition rates are reduced by energetic barriers, so flux is maximized by removing barriers and reducing the number of metastable states.
However, a greater number of metastable states, each representing a persistent conformation or ligand binding status of a biomolecular machine, make possible a larger array of schemes for 
the following: 
interaction, such as distinct binding affinities~\cite{mason99};
machine operation, such as `gating'~\cite{dogan15};
and 
regulation through variable action on distinct states~\cite{huse02}.

Multiple metastable states can be maintained by constraining the free energy landscape to preserve barriers. We implement these constraints by fixing the free energy differences between wells and either the barriers immediately before or immediately after them, equivalent to fixing the rate constants between discrete states for either the forward or reverse transitions. A \emph{\powerstrokenew} (\powerstrokenewacr) scheme (case A in~\cite{thomas01}) keeps the reverse free energy differences fixed, with the dissipation $\DISS_{ij}$ only modifying the forward rate constant,
\begin{equation}
\label{eq:psscheme}
\RCF = \BARE e^{\DISS_{ij}} \ \ \text{and} \ \  \RCR = \BARE \ ;
\end{equation}
whereas for a \emph{\brownianratchetnew} (\brownianratchetnewacr) scheme (case B in~\cite{thomas01}), the dissipation only modifies the reverse rate constant,
\begin{equation}
\label{eq:brscheme}
\RCF = \BARE \ \ \text{and} \ \ \RCR = \BARE e^{-\DISS_{ij}} \ .
\end{equation}
For the free energy landscape of Fig.~\ref{fig:Landscape}, the \powerstrokenewacr scheme has fixed $E^{\ddagger}_{12} - E_2$ and $E^{\ddagger}_{21} - (E_1 - \TOT)$, while the \brownianratchetnewacr scheme has fixed $E^{\ddagger}_{12} - E_1$ and $E^{\ddagger}_{21} - E_2$.

`Labile' here denotes the direction in which rate constants change with dissipation.  
This dependence of forward or reverse rate constants on dissipation for the \powerstrokenewacr and \brownianratchetnewacr schemes, respectively, is analogous to their dependence on the work performed by a motor in a `power stroke' or `Brownian ratchet'~\cite{wagoner16}.

\subsection{\PowerStrokenew Scheme}

\subsubsection{Two-state cycle}
\label{subsubsec:TwoStateFlux}
For a two-state \powerstrokenewacr cycle (Fig.~\ref{fig:Diagrams}a) with fixed total cycle dissipation $\TOT = \DISS_{12} + \DISS_{21}$, flux (Eq.~\eqref{eq:TwoStateFlux}) is maximized when the free energy dissipation of the first transition is
\begin{equation}
\label{eq:TwoStateFluxOpt}
\DOPT_{12} = \frac{1}{2}\DISS_{\text{tot}} + \frac{1}{2}\ln\frac{\BARETWOONE}{\BAREONETWO} \ .
\end{equation}
This produces equal forward rate constants,
\begin{equation}
\label{eq:TwoStateFluxRateConstants}
\RCF = \sqrt{\BAREONETWO \BARETWOONE}\ e^{\TOT/2} \ ,
\end{equation}
equal to the geometric mean of any full rate constants $\RCFONETWO$ and $\RCFTWOONE$ 
consistent with the bare rate constants $\BARE$ and total dissipation $\TOT$.

The optimal allocation of total dissipation $\TOT$ differs from the `naive' allocation $\frac{1}{2}\TOT$ that evenly divides the dissipation among the transitions. 
More specifically, the optimal deviation from the naive allocation is
$\DELD^*_{12} \equiv \DOPT_{12} - \frac{1}{2}\DISS_{\text{tot}} = \tfrac{1}{2}\ln(\BARETWOONE/\BAREONETWO)$.
The optimal allocation compensates for variation in bare rate constants, so for \powerstrokenewacr, transitions with larger $\BARE$ are optimally allocated less dissipation. In fact, the optimal allocation of dissipation to one transition is negative when 
$|\ln (\BARETWOONE/\BAREONETWO)| > \TOT$.

\subsubsection{Three-state cycle}
\begin{figure}[tbp] 
	\centering
	\hspace{0.0in}
	\begin{tabular}{c}
        \hspace{-0.100in}\includegraphics[width=3.375in]{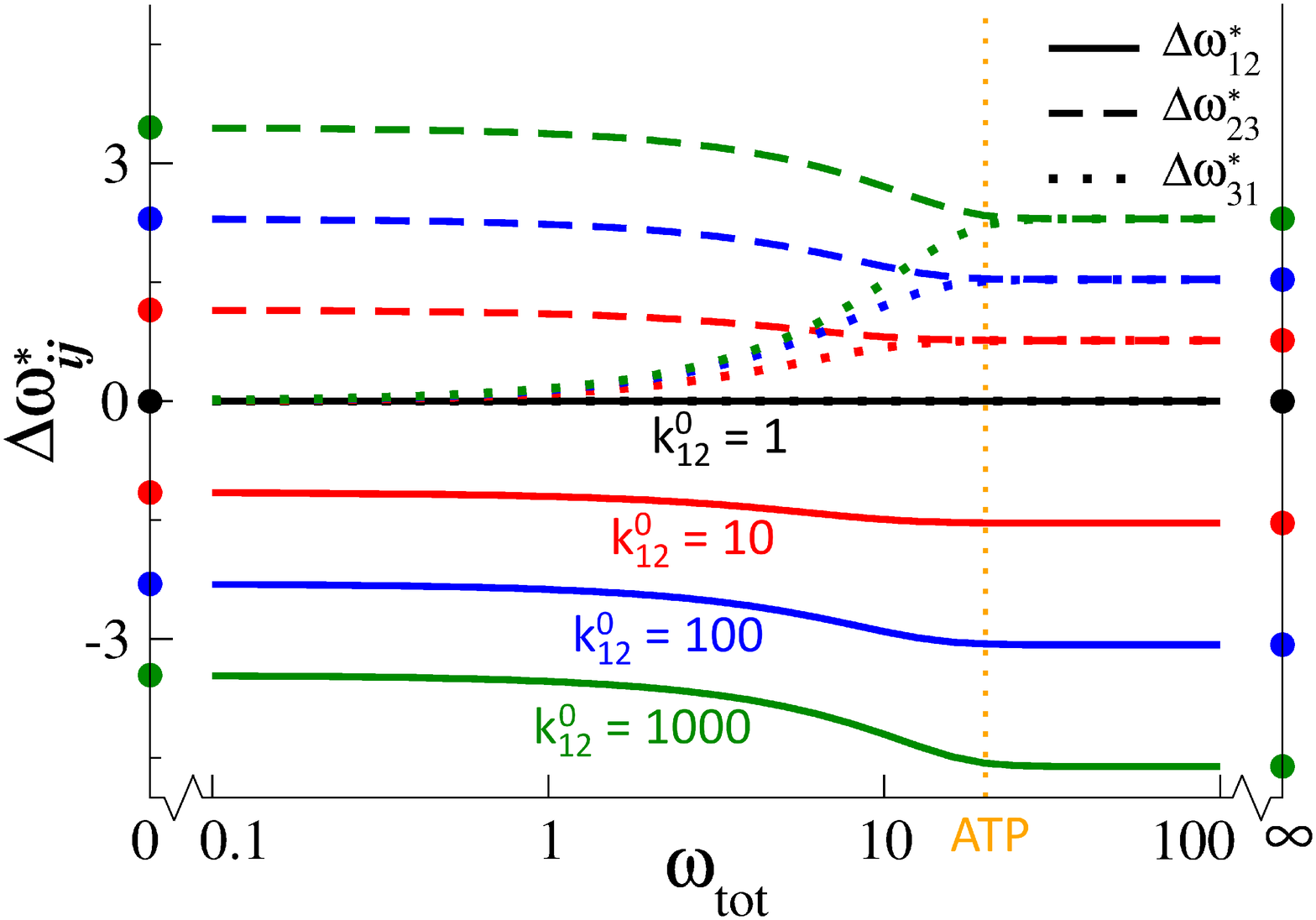}\\
	\end{tabular}
	\caption{\label{fig:ThreeStateFlux}
{\bf Uneven allocation of dissipation maximizes flux in a three-state \powerstrokenew cycle.} 
Dissipation allocations $\Delta\DOPT_{12}$ (solid curves), $\Delta\DOPT_{23}$ (dashed), and $\Delta\DOPT_{31}$ (dotted) for the three transitions in Fig.~\ref{fig:Diagrams}b that maximize the flux. Dissipations expressed as differences $\Delta\DOPT_{ij} \equiv \DOPT_{ij} - \frac{1}{3}\TOT$ from the naive allocation of equal dissipation $\tfrac{1}{3}\TOT$ to each transition. $\BARETWOTHREE = \BARETHREEONE = 1$ is fixed, and $\BAREONETWO$ varies with color. When $\BAREONETWO = 1$ (black), an even allocation of dissipation to each transition maximizes flux, so $\Delta\DOPT_{ij} = 0$ for all $\TOT$. As $\BAREONETWO$ increases (black$\to$red$\to$blue$\to$green), the flux-maximizing allocation increasingly deviates from an even allocation, as shown by the increasing magnitude of $\Delta\DOPT_{ij}$. Allocations at limiting $\TOT$ are shown by circles (low $\TOT$ plotted at $\TOT=0$, Eq.~\eqref{eq:HighEntropyProduction}; high $\TOT$ at $\TOT=\infty$, Eq.~\eqref{eq:LowDissipation}). Vertical dotted orange line at $\TOT=20$ represents the ATP hydrolysis free energy under physiological conditions.
}
\end{figure}

For a three-state cycle (Fig.~\ref{fig:Diagrams}b), the flux is~\cite{hill77}
\begin{equation}
\label{eq:ThreeStateFlux}
J = \dfrac{\RCFONETWO\RCFTWOTHREE\RCFTHREEONE - \RCRONETWO\RCRTWOTHREE\RCRTHREEONE}{\left(
\splitdfrac{\RCFONETWO\RCRTWOTHREE + 
\RCRONETWO\RCRTWOTHREE + \RCRTWOTHREE\RCRTHREEONE + \RCFONETWO\RCFTHREEONE + \RCRONETWO\RCFTHREEONE}
{+ \RCRONETWO\RCRTHREEONE + \RCFONETWO\RCFTWOTHREE + \RCFTWOTHREE\RCFTHREEONE + \RCFTWOTHREE\RCRTHREEONE}\right)} \ .
\end{equation}
For \powerstrokenewacr, Fig.~\ref{fig:ThreeStateFlux} shows the numerically determined allocation of free energy dissipation that maximizes the flux subject to a fixed $\DISS_{\text{tot}} = \DISS_{12} + \DISS_{23} + \DISS_{31}$ and fixed second and third bare rate constants $\BARETWOTHREE=\BARETHREEONE=1$, for several different $\BAREONETWO$ across multiple orders of magnitude~\cite{hwang17}. When $\BAREONETWO=1$, the $\DOPT_{ij}$ all equal the naive value $\frac{1}{3}\TOT$, as expected by symmetry. As $\BAREONETWO$ increases, the optimal allocations $\DOPT_{ij}$ depart from the naive case, with the dissipation $\DOPT_{12}$ of the first transition decreasing, and that of the second and third transitions ($\DOPT_{23}$ and $\DOPT_{31}$) increasing.

At high $\TOT$, the reverse flux is much smaller than the forward flux, $J_{ij-} = \RCR P_j \ll \RCF P_i = J_{ij+}$, and hence the net flux roughly equals the forward flux, $J = J_{ij+} - J_{ij-} \simeq J_{ij+}$.
In this limit, the cycle effectively only has forward transitions, leading (see Appendix) to an optimal dissipation allocation
\begin{equation}
\label{eq:HighEntropyProduction}
\DELD^*_{12} = \frac{1}{3}\ln\frac{\BARETWOTHREE\BARETHREEONE}{\left(\BAREONETWO\right)^2}
\end{equation}
that is independent of $\TOT$. Similar expressions for $\DELD_{23}^{*}$ and $\DELD_{31}^{*}$ are found by cyclically permuting the indices in Eq.~\eqref{eq:HighEntropyProduction}. These asymptotic values (Fig.~\ref{fig:ThreeStateFlux}, circles on the right edge) 
are indistinguishable from the 
limits of the 
numerical calculations. This optimal dissipation allocation produces equal forward rate constants 
\begin{equation}
\RCF = \BARE e^{\DISS_{ij}} = (\BAREONETWO\BARETWOTHREE\BARETHREEONE e^{\TOT})^{1/3} \ ,
\end{equation}
that are the geometric mean of any full rate constants $\RCFONETWO$, $\RCFTWOTHREE$, and $\RCFTHREEONE$, consistent with $\BARE$ and $\TOT$.

At low $\TOT$, when the net flux $J = J_{ij+} - J_{ij-}$ is much smaller than either the forward or reverse fluxes, $J \ll J_{ij+}$ and $J \ll J_{ij-}$, the optimal $\DISS_{ij}$ also asymptotically approach (generally nonzero) values independent of $\TOT$. Maximizing the net flux in Eq.~\eqref{eq:ThreeStateFlux} at low $\TOT$ leads (see Appendix) to an optimal allocation
\begin{equation}
\label{eq:LowDissipation}
\DISS_{12}^{*} = \frac{1}{2}\ln\frac{\BARETHREEONE}{\BAREONETWO} \ .
\end{equation}
Similar expressions for $\DOPT_{23}$ and $\DOPT_{31}$ are found by cyclically permuting the indices in Eq.~\eqref{eq:LowDissipation}. These asymptotic values (Fig.~\ref{fig:ThreeStateFlux}, circles on the left edge) show excellent agreement with the limiting numerical calculations.

At high total dissipation $\TOT$, the optimal allocation $\DOPT_{ij}$ reaches a limit where for all transitions the forward rates are much larger than the reverse rates, becoming effectively irreversible. For the three-state cycle, the 20$k_{\text{B}}T$ of free energy provided by ATP hydrolysis is near this limit (Fig.~\ref{fig:ThreeStateFlux}).  However, not all transitions will be effectively irreversible for a smaller dissipation budget per cycle step, which is obtained for machines that perform work against a resistive load or machines with more states per cycle.

\begin{figure}[tbp] 
	\centering
	\hspace{-0.0in}
	\begin{tabular}{c}
		\hspace{0.000in}\includegraphics[width=3.375in]{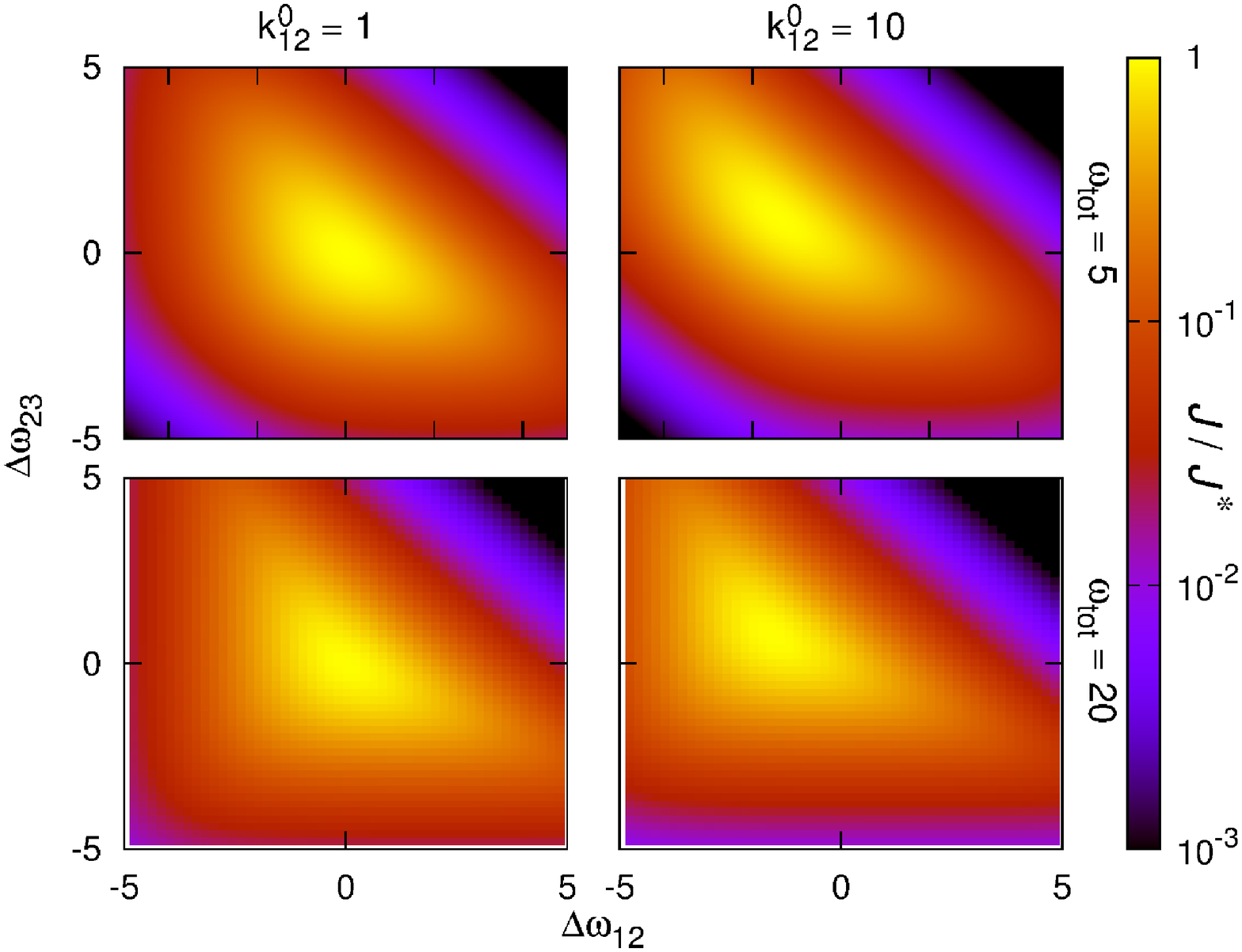}\\
	\end{tabular}
	\caption{\label{fig:Fluxes} 
{\bf Flux is sensitive to dissipation allocation.} Flux ratio $J/J^*$ as a function of the dissipation allocation, for the three-state \powerstrokenew cycle with $\BARETWOTHREE = \BARETHREEONE = 1$, and $\BAREONETWO$ and $\TOT$ varying across subplots. Dissipation allocations are shown as differences $\DELD_{ij} \equiv \DISS_{ij} - \tfrac{1}{3}\TOT$ from naive values $\tfrac{1}{3}\TOT$. Optimal flux $J^*$ is specific to each subplot.}
\end{figure}

The results above demonstrate that the optimal allocation of dissipation can significantly differ from an equal allocation to each transition. Fig.~\ref{fig:Fluxes} shows the variation of flux as the dissipation allocation is varied away from the optimal allocation. Exploring a range of several $k_{\rm B}T$ around the optimal allocation, the flux varies by more than three orders of magnitude. Thus a dissipation allocation significantly different from the optimal one can qualitatively alter the cycle output.

\subsection{\BrownianRatchetnew Scheme}
\label{subsec:PSBR}

For a two-state \brownianratchetnewacr cycle (as opposed to a \powerstrokenewacr cycle), the flux-maximizing allocation of dissipation is (Appendix)
\begin{equation}
\label{eq:TwoStateOpt}
\DELD^*_{12} = -\frac{1}{2}\ln\frac{\BARETWOONE}{\BAREONETWO} \ .
\end{equation}
The deviations in Eq.~\eqref{eq:TwoStateOpt} from naive allocations are identical to the \powerstrokenewacr result (Eq.~\eqref{eq:TwoStateFluxOpt}), except assigned to the other transition. 
Despite this apparent difference, in each state the probability that the next transition will move in the forward or reverse direction (and hence the ratio of one-sided fluxes) is identical for the optimized \powerstrokenewacr and \brownianratchetnewacr cycles.
For example, both schemes produce one-sided fluxes departing from state 1 that satisfy (see Appendix)
\begin{equation}
\label{eq:sameratio}
\frac{J^*_{12+}}{J^*_{21-}} = \sqrt{\frac{\BAREONETWO}{\BARETWOONE}}e^{\TOT/2} \ .
\end{equation}

\begin{figure}[h] 
	\centering
	\hspace{-0.0in}
	\begin{tabular}{c}
		\hspace{-0.300in}\includegraphics[width=3.375in]{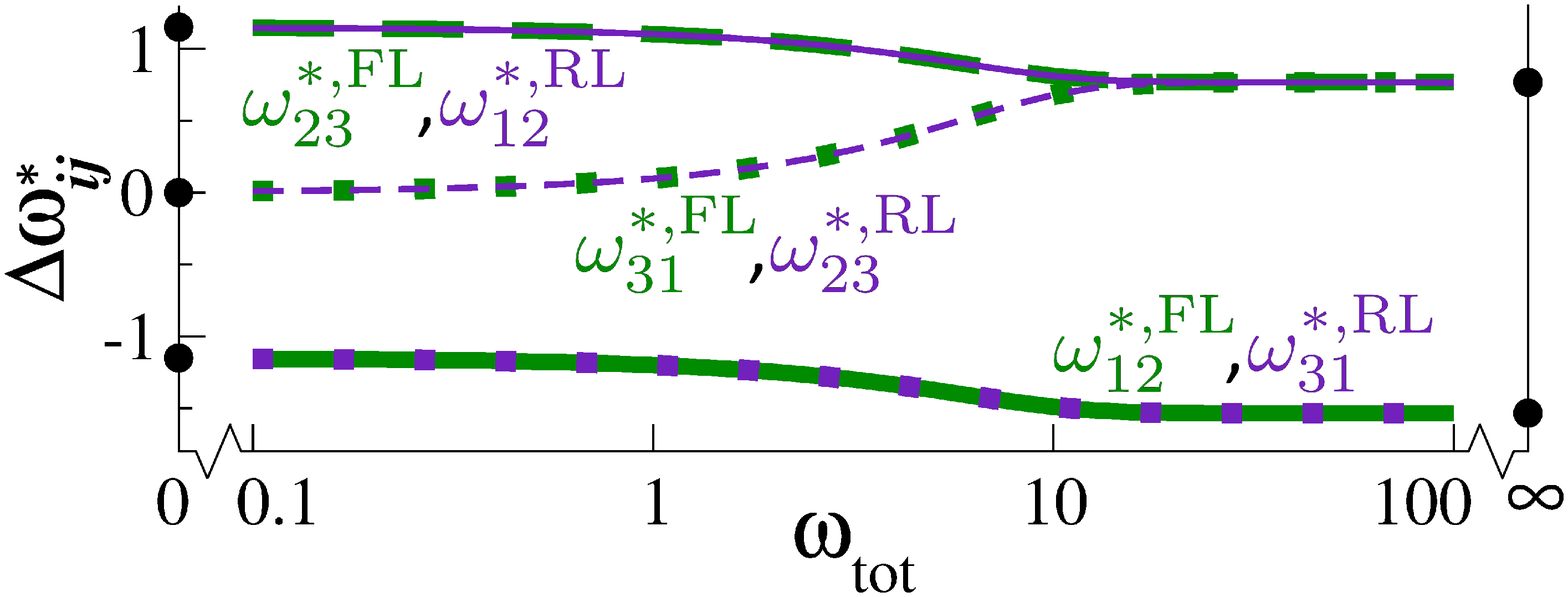}\\
        \hspace{-0.000in}\includegraphics[width=3.375in]{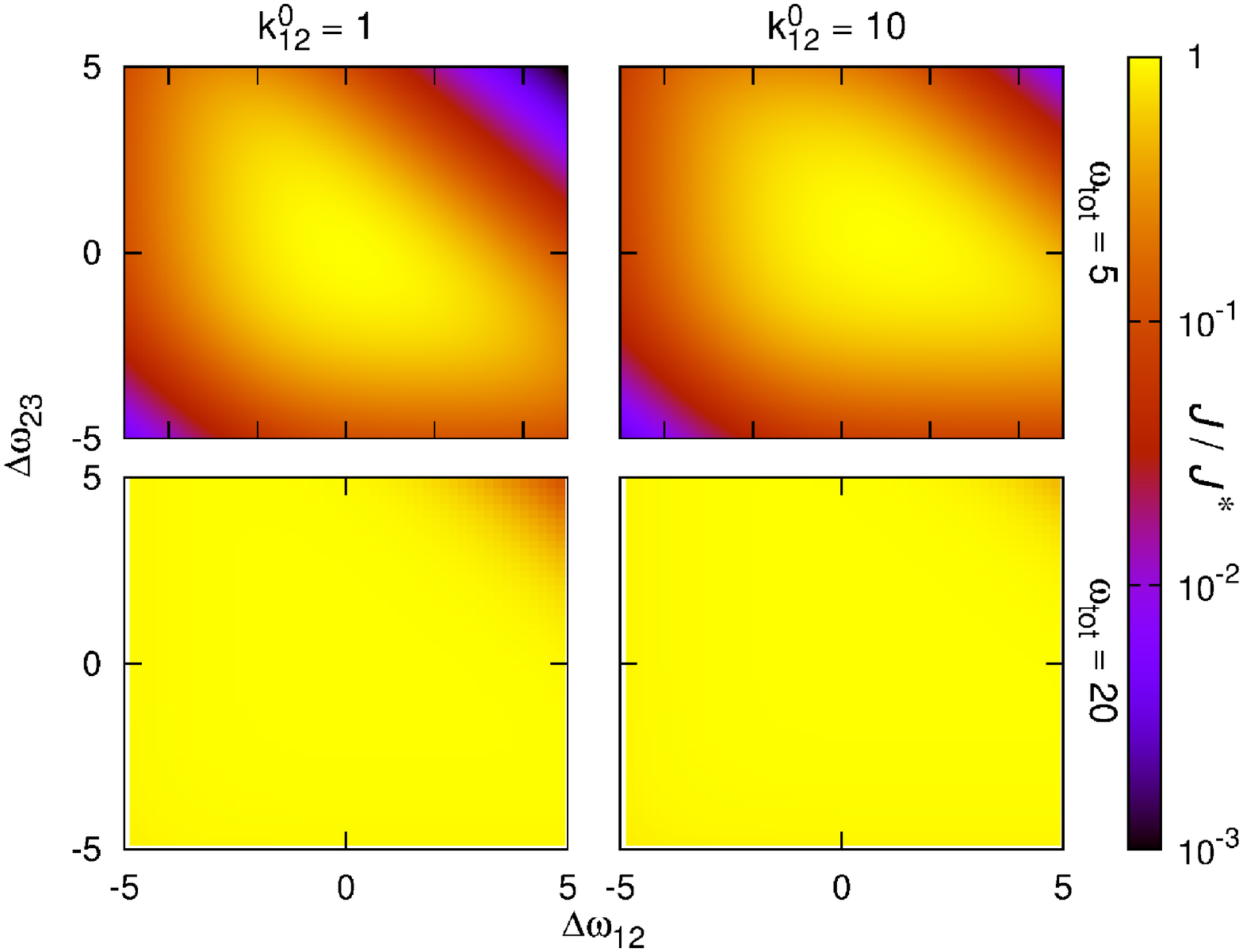}\\
	\end{tabular}
	\caption{\label{fig:PsBr} 
{\bf 
Optimal and sub-optimal allocations of dissipation for \brownianratchetnew scheme.}
(a) Allocation of dissipation $\DISS_{ij}$ to maximize flux around the three-state cycle, for \powerstrokenew (green) and \brownianratchetnew (purple) schemes.  
The optimal dissipation values are identical for the two mechanisms (as illustrated by the overlapping green and purple curves on the plot), however the individual dissipations are allocated to different transitions. Circles on the left and right edges show limiting optimal allocations at low and high total dissipation $\TOT$ (Eq.~\eqref{eq:BR_B} and Eq.~\eqref{eq:BR_A}), respectively. $\BAREONETWO = 10$ and $\BARETWOTHREE = \BARETHREEONE = 1$. (b) Flux sensitivity to dissipation allocation, analogous to Fig.~\ref{fig:Fluxes}, except for \brownianratchetnew scheme instead of \powerstrokenew.
}
\end{figure}

Similarly, for the three-state cycle, both mechanisms allocate dissipation identically, except cyclically permuted (Fig.~\ref{fig:PsBr}a). Thus optimal dissipation allocation in the three-state cycle also produces one-sided flux ratios that do not depend on the mechanism. The limiting optimal allocations at high $\TOT$ are (Appendix) 
\begin{equation}
\label{eq:BR_A}
\DELD_{12}^* = \frac{1}{3}\ln\frac{\BAREONETWO \BARETHREEONE}{\left(\BARETWOTHREE\right)^2} \ ,
\end{equation}
and at low $\TOT$ are (Appendix)
\begin{equation}
\label{eq:BR_B}
\DOPT_{12} = \frac{1}{2}\ln\frac{\BAREONETWO}{\BARETWOTHREE} \ .
\end{equation}

These results are 
intuitive: 
an \brownianratchetnewacr scheme can adjust reverse but not forward rates, and so to maximize the flux it allocates more dissipation to decelerate the fastest reverse rates (those with high $\BARE$); whereas an \powerstrokenewacr scheme can adjust forward but not reverse rates, so it allocates more dissipation to accelerate the slowest forward rates (with low $\BARE$).

Figure~\ref{fig:PsBr}b shows the dependence of three-state \brownianratchetnewacr flux on the dissipation allocation. For low $\TOT$ the flux varies substantially (by orders of magnitude) across allocations that differ by a few $k_{\rm B}T$ (similar to \powerstrokenewacr in Fig.~\ref{fig:Fluxes}), but at high $\TOT$ there is little variation of flux with dissipation allocation. The \brownianratchetnewacr flux is less sensitive to the dissipation allocation at high $\TOT$ because once reverse rates are sufficiently suppressed to be negligible (\emph{i.e.},\ for $e^{-\TOT/3} \ll 1$), reallocation of dissipation has reduced effect.  

For given bare rates $\BARE$ and total dissipation $\TOT$, an \powerstrokenewacr cycle will always produce more flux than the corresponding \brownianratchetnewacr cycle, similar to previous results~\cite{wagoner16}.
\powerstrokenewacr and \brownianratchetnewacr schemes represent extremes of a more general mechanism, whereby some dissipation is spent speeding up the forward transitions (as for \powerstrokenewacr), and the remaining fraction slows down the reverse transitions (as for \brownianratchetnewacr):
\begin{equation}
\label{eq:both}
\RCF = \BARE e^{\DISS_{ij}^+} \ \ \text{and} \ \  \RCR = \BARE e^{-\DISS_{ij}^-} \ .
\end{equation}
This is similar to splitting force-dependence among reaction rates in previous studies~\cite{fisher99,wagoner16}. $\sum (\DISS_{ij}^+ + \DISS_{ij}^-) = \TOT$ is fixed, leaving $2n-1$ free parameters to optimize over in an $n$-state cycle. For any given dissipation allocation, flux can always be increased by shifting some dissipation $\delta\DISS$ from slowing the reverse rate to speeding the corresponding forward rate. This is equivalent to simply lowering the barriers for the free energy landscape in Fig.~\ref{fig:Landscape}, which removes the distinction between states, as discussed above.

\section{Discussion}
The Second Law of thermodynamics requires free energy dissipation to break detailed balance and maintain directed flux
~\cite{machta15},
but does not specify a quantitative relationship between dissipation and flux~\cite{barato15,feng08,brown16}. 

From the perspective of a free energy landscape, 
we find that the flux is increased by lowering the barriers so that they are no longer effective. This is intuitive, as transition rates are reduced by energetic barriers, and suggests that molecular machines should reduce the number of metastable states to increase forward flux. However, molecular machines perform their tasks using multiple metastable states, and accordingly we have focused on scenarios that allow distinct states to be maintained.

In a reaction cycle with a fixed number of discrete states, we have shown that flux is maximized by an uneven allocation of a fixed dissipation budget among the various discrete transitions, compensating for differences in the bare rate constants of each transition (see Eq.~\eqref{eq:TwoStateFluxOpt} and Eq.~\eqref{eq:TwoStateOpt}, Figs.~\ref{fig:ThreeStateFlux} and \ref{fig:PsBr}a). This is related to recent findings that flux is affected differently by adjusting the bare rate of different transitions~\cite{wagoner16}. The flux can be quite sensitive to the precise dissipation allocation (Figs.~\ref{fig:Fluxes} and \ref{fig:PsBr}b), suggesting a significant cost to non-optimal allocations.

This result differs from the uniform allocations found to be optimal in various other contexts, including maximizing power at fixed entropy production rate \cite{qian16}, minimizing entropy production at fixed flux~\cite{qian00, sauar96, johannessen05}, maximizing free energy conversion efficiency~\cite{oster00}, and minimizing the dissipation cost of a given precision~\cite{barato15}. 
Several other studies have argued that to maintain a high flux, large free energy increases should be broken up into smaller pieces, with no individual free energy change too large~\cite{hill81, yu07, anandakrishnan16}. Even in synthetic molecular motors, it is thought that similar forward rates are optimal (to avoid `traffic jams')~\cite{geertsema09}.

We find an unequal optimal dissipation allocation occurs when: the nonequilibrium steady-state flux is maximized
; optimization is subject to fixed total dissipation budget per cycle
; the ratio of forward and reverse rate constants varies exponentially, not linearly, with dissipation (Eq.~\eqref{eq:ratio}); and cycle transitions have different bare rate constants, corresponding to different barrier heights and effective diffusivities. For some previous studies finding even dissipation allocations to be optimal, a single change is sufficient to make uneven allocations optimal: \emph{e.g.}, imposing distinct bare rate constants~\cite{barato15} or changing the dependence of flux on dissipation from linear (the near-equilibrium case) to exponential~\cite{sauar96}.

Many models parametrized to biomolecular machine dynamics contain effectively irreversible transitions, \emph{e.g.}, models of kinesin~\cite{clancy11,visscher99}, myosin~\cite{cappello07, xie06}, RNA polymerase~\cite{abbondanzieri05}, and viral packaging motors~\cite{moffitt09,liu14}. Such irreversible transitions are, strictly speaking, unphysical due to their violation of microscopic reversibility~\cite{fisher99, astumian15}; in reality they represent a forward rate constant much larger than the reverse rate constant, a signature of large dissipation over that transition. Since other transitions in these models are reversible, this implies the dissipation allocation in such models must be highly unequal, consistent with the uneven dissipation allocation we find maximizes flux.

Other models of driven biomolecular cycles---such as in myosin~\cite{fox98,skau06,wu07} and several enzymes~\cite{hwang17}---lack explicitly irreversible transitions, but have ratios of forward and reverse rate constants, and hence free energy dissipation, that vary significantly across the different reactions composing a cycle. 

Dissipation biases forward and reverse rate constants, but there is no unique way to achieve this bias~\cite{thomas01,wagoner16}. We explored in detail two extremes for how dissipation can lead to biased progress: a \powerstrokenew scheme, where dissipation increases forward rate constants; and a \brownianratchetnew scheme, where dissipation decreases reverse rate constants. Although \powerstrokenewacr and \brownianratchetnewacr mechanisms lead to distinct optimal allocations of dissipation, both lead to identical transition probability ratios from each state (Eq.~\eqref{eq:sameratio}). An \powerstrokenewacr cycle produces more flux than a comparable \brownianratchetnewacr cycle, but \powerstrokenewacr flux is quite sensitive to the dissipation allocation (Fig.~\ref{fig:Fluxes}), while \brownianratchetnewacr flux is insensitive to the dissipation allocation for a large free energy budget (Fig.~\ref{fig:PsBr}b).

On evolutionary timescales, mutations alter the conformational free energies of initial and final states differently.  For a transition state conformationally similar to the initial state, a mutation should produce similar changes in the initial and transition state free energies, so the forward rate should change less than the backward rate. This is analogous to the distance to the transition state affecting the sensitivity of unfolding rates to applied force~\cite{elms12}. Our \powerstrokenewacr and \brownianratchetnewacr mechanisms thus correspond to a transition state conformationally similar to the final and to the initial states, respectively (see Appendix, Fig.~\ref{fig:NearFar}).

\begin{figure}[tbp] 
	\centering
	\hspace{-0.0in}
	\begin{tabular}{c}
		\hspace{0.000in}\includegraphics[width=3.375in]{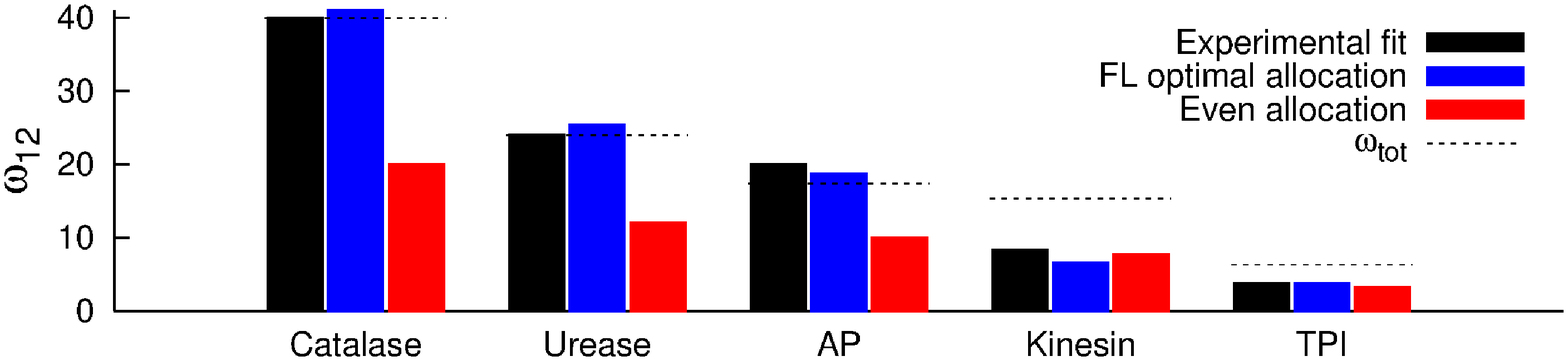} \\
	\end{tabular}
	\caption{\label{fig:Comparison}
{\bf Forward labile predictions better match experimental dissipation allocations than do even allocations.}
Dissipation of transition $\DISS_{12}$ in several enzymes, from fit to experiment (black)~\cite{hwang17}, flux optimization under the \powerstrokenewacr scheme (blue), and even allocation (red). AP, alkaline phosphatase; TPI, triose phosphate isomerase. Details in Appendix.
}
\end{figure}

Optimizing an \powerstrokenewacr cycle predicts that transitions with low bare 
rate constants will be allocated more dissipation, while for an \brownianratchetnewacr cycle high bare rate constants are allocated more dissipation. Dissipation allocations for several two-state enzyme models~\cite{hwang17} fit to experiment are closely matched by the FL optimal allocation, generally much better than by an even allocation (Fig.~\ref{fig:Comparison}) or the RL optimal allocation. See Appendix for more extensive comparisons.

We expect that adjusting the dissipation allocation (of a fixed total dissipation
per cycle) would require only isolated changes in molecular machine dynamics, primarily affecting machine output or productivity, and minimally impacting the rest of the cell.  In contrast, adjusting the dissipation budget, for example through the free energy of ATP hydrolysis, would affect numerous driven processes throughout the cell.

Adjustment of the dissipation allocation through isolated mutations is supported by experimental findings. Point mutations in the kinesin-1 nucleotide binding pocket likely affect the dissipation allocation by altering the size of the pocket~\cite{kapoor99} or the ADP unbinding rate~\cite{higuchi04,uchimura10} and lead to significant decreases in kinesin velocity or ATP hydrolysis rate while remaining functional. Changes in binding affinity due to mutation (e.g., 2.5-fold change for a transcription regulator~\cite{huang16}, or 40-fold for a membrane regulatory protein~\cite{feig88}) correspond to a different unbinding rate, which also would change the dissipation allocation.

Our optimizations omit several significant biophysical considerations. 
For example, we allow rate constants to vary without bound, although practically they are limited by molecular diffusion. We also focus on a single biomolecular cycle; an interesting extension would be to investigate the effects of alternative pathways thought to be present in biomolecular machines~\cite{clancy11,wu07}. Further elaborations of this work could explore the sensitivity of flux to (varying) resistive forces as well as cycle states vulnerable to `escape' (such as a molecular motor falling off its track).

\begin{acknowledgments}
This work was supported by a Natural Sciences and Engineering Research Council of Canada (NSERC) Discovery Grant, a Tier II Canada Research Chair, and by funds provided by the Faculty of Science, Simon Fraser University through the President's Research Start-up Grant, and was enabled in part by support provided by WestGrid and Compute Canada Calcul Canada. The authors thank N.\ Forde, J.\ Bechhoefer, E.\ Emberly, N.\ Babcock, S.\ Large, A.\ Kasper, E.\ Lathouwers, S.\ Blaber, J.\ Lucero (SFU Physics), P. Unrau, D. Sen (SFU Molecular Biology and Biochemistry), A. Bennet (SFU Chemistry), T.\ Shitara (University of Tokyo), and W.\ Hwang (Korea Institute for Advanced Study) for useful discussions and feedback.
\end{acknowledgments}

\appendix

\section{Optimizations with varying number of metastable states}
\label{sec:landscape_app}

We investigate the free energy landscape of Fig.~\ref{fig:Landscape}.
For simplicity we set $E_1 - \TOT = 0$ and hence $E_1 = \TOT$. The rate constants $k_{12}^{+/-}$ for forward and reverse transitions over the first barrier, with free energy $E^{\ddagger}_{12}$, are, respectively
\begin{equation}
\label{eq:landscape_app_rates1}
\RCFONETWO = \tau^{-1}_{12}e^{-(E^{\ddagger}_{12} - \TOT)} \ \ \text{and} \ \  \RCRONETWO = \tau^{-1}_{12}e^{-(E^{\ddagger}_{12} - E_2)} \ .
\end{equation}
The forward and reverse rate constants over the second barrier, with free energy $E^{\ddagger}_{21}$, are
\begin{equation}
\label{eq:landscape_app_rates2}
\RCFTWOONE = \tau^{-1}_{21}e^{-(E^{\ddagger}_{21} - E_2)} \ \ \text{and} \ \  \RCRTWOONE = \tau^{-1}_{21}e^{-E^{\ddagger}_{21}} \ .
\end{equation}
The steady-state flux for a two-state cycle is~\cite{hill77}
\begin{equation}
\label{eq:TwoStateFluxApp}
J = \frac{\RCFONETWO\RCFTWOONE - \RCRONETWO\RCRTWOONE}{\RCFONETWO + \RCRONETWO + \RCFTWOONE + \RCRTWOONE} \ .
\end{equation}
Inserting Eqs.~\eqref{eq:landscape_app_rates1} and \eqref{eq:landscape_app_rates2} into \eqref{eq:TwoStateFluxApp} and rearranging gives
\begin{equation}
\label{eq:FluxWithRates}
J = \frac{e^{\TOT} - 1}{\tau_{12}e^{E^{\ddagger}_{12}}(1 + e^{-E_2}) + \tau_{21}e^{E^{\ddagger}_{21}}(1 + e^{\TOT - E_2})} \ .
\end{equation}
We consider the states at $E=\TOT$ and $E=0$ to be fixed, as varying them relative to one another changes the free energy budget $\TOT$. We vary $E_2$, $E^{\ddagger}_{12}$, and $E^{\ddagger}_{21}$ to maximize the flux $J$. Assuming barriers are higher than states, straightforward differentiation shows that
\begin{equation}
\label{eq:landscape_app_inequalities1}
\frac{\partial J}{\partial E^{\ddagger}_{12}} < 0 \ \ \text{and} \ \ \frac{\partial J}{\partial E^{\ddagger}_{21}} < 0 \ ,
\end{equation}
\emph{i.e.}, flux increases as either barrier height decreases. Once the barrier energies are at or below the neighboring state energies, Eqs.~\eqref{eq:landscape_app_rates1} and \eqref{eq:landscape_app_rates2} no longer hold.
Eq.~\eqref{eq:landscape_app_inequalities1} indicates that the flux is increased by removing the barriers altogether, leaving state 2 (at energy $E_2$) no longer metastable.

In a separate optimization, 
we constrain the barriers at fixed energies $E^{\ddagger}_{12}$ and $E^{\ddagger}_{21}$, and then vary $E_2$. Differentiating Eq.~\eqref{eq:FluxWithRates}, again subject to barriers higher than states, gives
\begin{equation}
\frac{\partial J}{\partial E_2} > 0 \ ,
\end{equation}
meaning that the flux increases as $E_2$ increases.

The previous two optimizations allowed either the barrier energies to decrease below the neighboring state energies, or $E_2$ to rise above the barrier energies. 
We now consider a scenario where the free energy differences $\Delta E_{12}^- = E^{\ddagger}_{12} - E_2$ and $\Delta E_{21}^+ = E^{\ddagger}_{21} - E_2$ are fixed, so that $E_2$, $E^{\ddagger}_{12}$, and $E^{\ddagger}_{21}$ move up and down together. This gives rate constants
\begin{subequations}
\label{eq:landscape_app_rates3}\begin{align}
\RCFONETWO &= \tau^{-1}_{12}e^{-(E_2 + \Delta E_{12}^- - \TOT)} \\ 
\RCRONETWO &= \tau^{-1}_{12}e^{-\Delta E_{12}^-} \\
\RCFTWOONE &= \tau^{-1}_{21}e^{-\Delta E_{21}^+} \\ 
\RCRTWOONE &= \tau^{-1}_{21}e^{-(E_2 + \Delta E_{21}^+)} \ .
\end{align}
\end{subequations}
Substituting these rate constants into Eq.~\eqref{eq:TwoStateFluxApp} gives
\begin{equation}
J = \frac{e^{\TOT} - 1}{e^{E_2}(\tau_{12}e^{\Delta E_{12}^-} + \tau_{21}e^{\Delta E_{21}^+}) + \tau_{12}e^{\Delta E_{12}^-} + \tau_{21}e^{\TOT + \Delta E_{21}^+} } \ .
\end{equation}
When barriers are higher than states,
\begin{equation}
\frac{\partial J}{\partial E_2} < 0 \ ,
\end{equation}
meaning the flux increases as $E_2$ decreases. This continues until one of the barriers is at or below one of the other two states, when Eq.~\eqref{eq:landscape_app_rates3} no longer holds. This optimization, similar to the previous two optimizations, increases the flux by removing the effect of the barriers.

\begin{figure}[tbp] 
	\centering
	\hspace{-0.0in}
	\begin{tabular}{cc}
		\hspace{-0.300in}\includegraphics[width=2.5in]{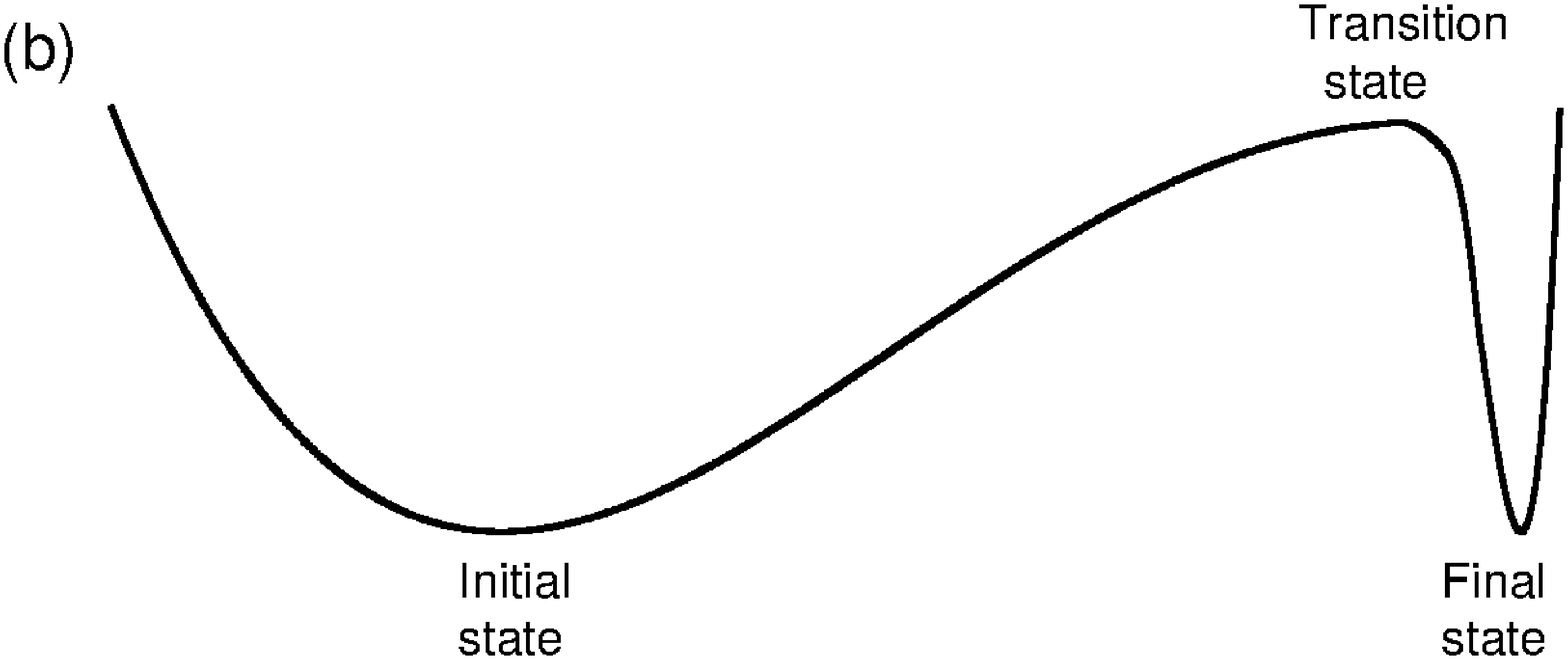}\\
        \hspace{-0.300in}\includegraphics[width=2.5in]{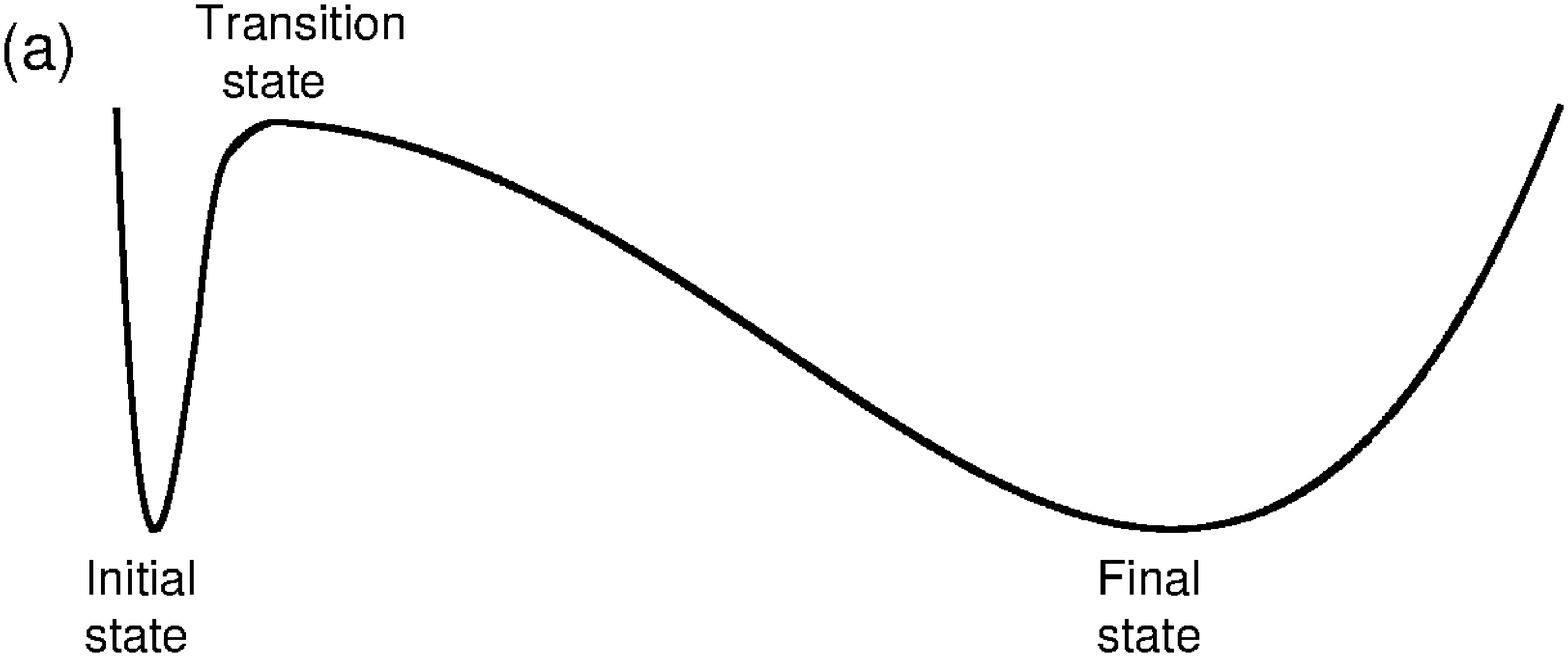}
	\end{tabular}
	\caption{\label{fig:NearFar} 
{\bf Correspondence between energy landscapes and forward and reverse labile schemes.} 
Forward labile and reverse labile schemes use dissipation to change only forward and reverse rate constants, respectively. In analogy to force-induced unfolding~\cite{elms12}, the proximity of the transition state can lead to forward and reverse rates differing in their sensitivity to dissipation. (a) The forward labile scheme corresponds to a transition state quite close to the final state. Changes in the free energy difference between initial and final state (dissipation) lead to a change in the difference between initial and transition states, but not a significant change in the difference between the final and transition states. This changes the forward rate, but not the reverse. (b) Conversely, the reverse labile scheme corresponds to a transition state quite close to the initial state. Dissipation changes lead to relative changes between the final and transition states, but not the initial and transition states. This changes the reverse rate, but not the forward. These two scenarios are extremes; the transition state could be anywhere between the initial and final states.
}
\end{figure}

\section{Additional model details}

We describe our cycles with `basic' free energy differences $\DISS_{ij}$~\cite{hill81,hill83}, because they directly relate to the ratio of forward and reverse transition rate constants in Eq.~\eqref{eq:ratio}.
Unlike basic free energy differences, `gross' free energy changes also include the entropic contribution associated with transitions between states with different occupation probabilities~\cite{hill83}. At steady state, the entropic contributions included in the gross free energy cancel out over a complete cycle, making the basic and gross free energy budgets identical.

\section{\PowerStrokenew Scheme}

\subsection{Two-state flux}
\label{subsec:TwoState_App}
For a two-state cycle, Eq.~\eqref{eq:TwoStateFluxApp} gives the steady-state flux.

Each transition has a bare rate constant $\BARE$. The log-ratio of the full rate constants is the dissipation $\DISS_{ij}$ of each transition, $\RCF/\RCR = e^{\DISS_{ij}}$. For a \powerstrokenew scheme, dissipation increases the forward rate constant, $\RCF = \BARE e^{\DISS_{ij}}$, and leaves unchanged the reverse rate constant $\RCR = \BARE$ (see main text). With these expressions, the flux can be rewritten as
\begin{equation}
\label{eq:FluxTwoStatePS}
J = \frac{\BAREONETWO\BARETWOONE\left(e^{\TOT}- 1\right)}
{\BAREONETWO \left(e^{\DISS_{12}} + 1\right) + \BARETWOONE \left(e^{\DISS_{21}} + 1\right)} \ .
\end{equation}

We consider reaction cycles with a fixed total free energy dissipation $\TOT = \DISS_{12} + \DISS_{21}$. The dissipation allocation is determined by the single free parameter $\DISS_{12}$, without loss of generality; the other transition's dissipation $\DISS_{21} = \TOT - \DISS_{12}$ is then fixed. Setting $\MD J/\MD\DISS_{12} = 0$ gives
\begin{equation}
\label{eq:twostatefluxoptapp}
\DOPT_{12} = \frac{1}{2}\TOT + \frac{1}{2}\ln\frac{\BARETWOONE}{\BAREONETWO} \ .
\end{equation}
The corresponding optimal flux is
\begin{equation}
\label{eq:FluxTwoStatePSOpt}
J^* = \frac{\BAREONETWO\BARETWOONE(e^{\TOT} - 1)}{\BAREONETWO + \BARETWOONE + 2\sqrt{\BAREONETWO\BARETWOONE}e^{\TOT/2}} \ .
\end{equation}

To quantify how $J/J^*$ decreases from 1 away from $\DOPT_{12}$, first we solve for $J/J^*$ near $\DOPT_{12}$.  Differentiating Eq.~\eqref{eq:FluxTwoStatePS} gives
\begin{equation}
\frac{\MD J}{\MD \DISS_{12}} = \frac{\BAREONETWO\BARETWOONE(e^{\TOT} - 1)(\BARETWOONE e^{\TOT-\DISS_{12}} - \BAREONETWO e^{\DISS_{12}})}{(\BAREONETWO + \BARETWOONE + \BAREONETWO e^{\DISS_{12}} + \BARETWOONE e^{\TOT - \DISS_{12}})^2} \ .
\end{equation}
For $\DISS_{12} = \DOPT_{12} + \delta$, expanding to first order in $\delta$ produces
\begin{equation}
\frac{\MD J}{\MD\DISS_{12}} \simeq -\left(\frac{1}{2}\frac{\BAREONETWO + \BARETWOONE}{\sqrt{\BAREONETWO\BARETWOONE}} e^{-\TOT/2} + 1\right)^{-1}\delta \ .
\end{equation}
Integrating and rearranging gives, for small $|\delta| = |\DISS_{12} - \DOPT_{12}|$,
\begin{equation}
\label{eq:FluxCloseApproxPS}
\frac{J}{J^*} \simeq 1 - \left(\frac{\BAREONETWO + \BARETWOONE}{\sqrt{\BAREONETWO\BARETWOONE}} e^{-\TOT/2} + 2\right)^{-1}
\delta^2 \ .
\end{equation}

\begin{figure}[tbp] 
	\centering
	\hspace{-0.0in}
	\begin{tabular}{c}
		\hspace{-0.300in}\includegraphics[width=3.375in]{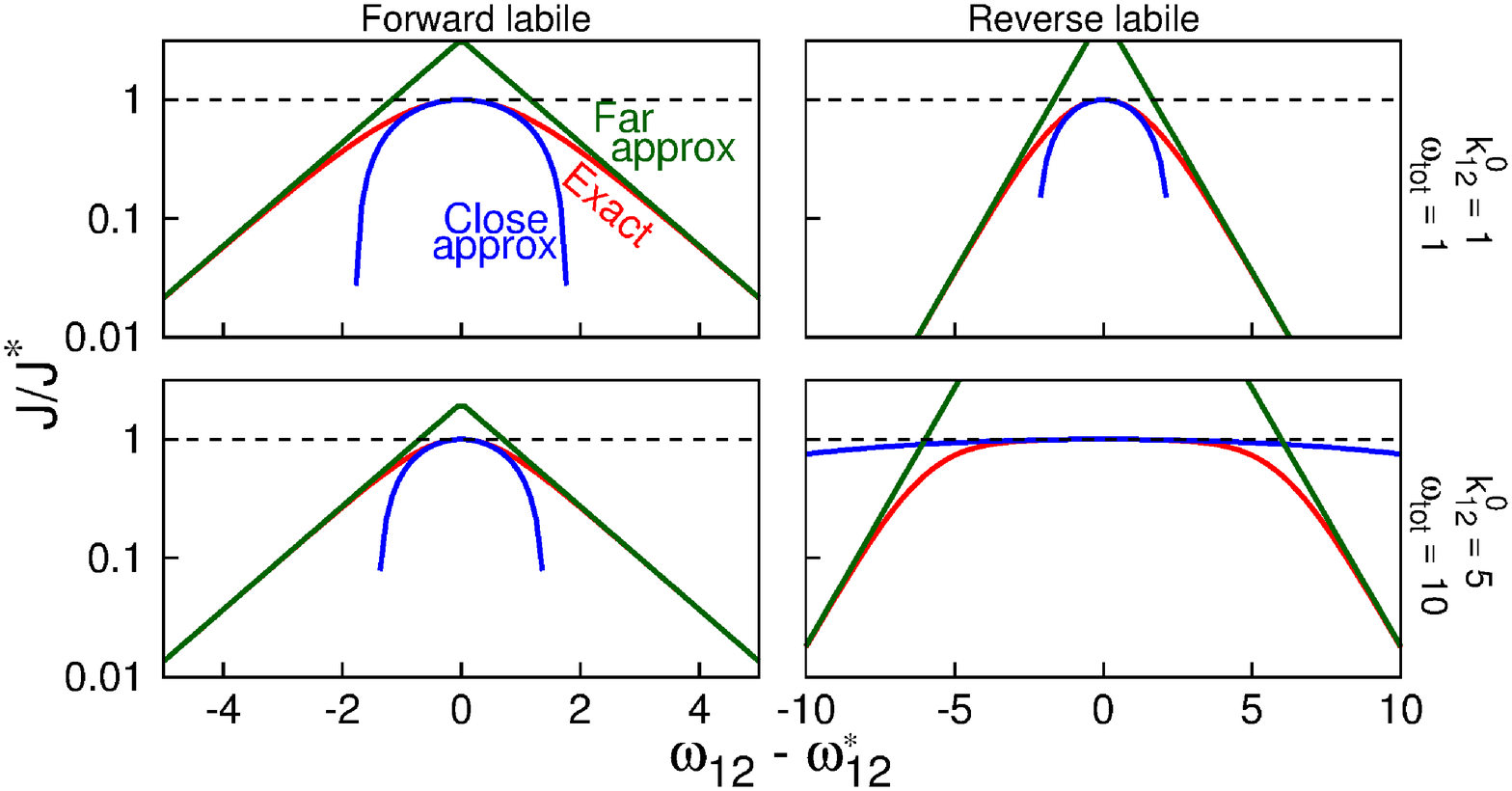}
	\end{tabular}
	\caption{\label{fig:TwoStateSensitivity} 
{\bf Two-state flux sensitivity.} 
Flux ratio $J/J^*$ as a function of the dissipation allocation, for the two-state cycle with $k_{21}^0 = 1$, and $k_{12}^0$ and $\TOT$ as indicated. For \powerstrokenew (\brownianratchetnew) cycles, $J/J^*$ is given by Eqs.~\eqref{eq:FluxTwoStatePS} and \eqref{eq:FluxTwoStatePSOpt} (\eqref{eq:FluxTwoStateBR} and \eqref{eq:FluxTwoStateBROpt}), the close approximation by Eq.~\eqref{eq:FluxCloseApproxPS} (\eqref{eq:FluxCloseApproxBR}), and the far approximation by Eq.~\eqref{eq:FluxFarApproxPS} (\eqref{eq:FluxFarApproxBR}).  Dashed black line indicates $J/J^* = 1$.
}
\end{figure}

For $J/J^*$ far from $\DOPT_{12}$, we divide Eq.~\eqref{eq:FluxTwoStatePS} by Eq.~\eqref{eq:FluxTwoStatePSOpt},
\begin{equation}
\frac{J}{J^*} = \frac{\BAREONETWO + \BARETWOONE + 2\sqrt{\BAREONETWO\BARETWOONE}e^{\TOT/2}}{\BAREONETWO + \BARETWOONE + \BAREONETWO e^{\DISS_{12}} + \BARETWOONE e^{\TOT - \DISS_{12}}} \ .
\end{equation}
Rewriting $\BAREONETWO e^{\DISS_{12}} = \sqrt{\BAREONETWO\BARETWOONE}e^{\TOT/2}e^{\delta}$ and $\BARETWOONE e^{\TOT - \DISS_{12}} = \sqrt{\BAREONETWO\BARETWOONE}e^{\TOT/2}e^{-\delta}$ gives
\begin{equation}
\frac{J}{J^*} = \frac{\BAREONETWO + \BARETWOONE + 2\sqrt{\BAREONETWO\BARETWOONE}e^{\TOT/2}}{\BAREONETWO + \BARETWOONE + \sqrt{\BAREONETWO\BARETWOONE}e^{\TOT/2}(e^{\delta} + e^{-\delta})} \ .
\end{equation}
For large $|\delta| = |\DISS_{12} - \DOPT_{12}|$,
\begin{equation}
\label{eq:FluxFarApproxPS}
\frac{J}{J^*} \simeq \left(\frac{\BAREONETWO + \BARETWOONE}{\sqrt{\BAREONETWO\BARETWOONE}} e^{-\TOT/2} + 2\right)e^{-|\delta|} \ .
\end{equation}  
Fig.~\ref{fig:TwoStateSensitivity} compares Eqs.~\eqref{eq:FluxCloseApproxPS} and \eqref{eq:FluxFarApproxPS} to the exact $J/J^*$, showing good agreement in the expected regimes.

\subsection{Three-state flux for high $\DISS_{\text{tot}}$}
\label{subsec:ThreeStateHigh_App}

For high total dissipation $\TOT$, the forward rate constants are exponentially increased, and the backward rate constants are negligible in comparison, 
producing a cycle with effectively only forward rates. A three-state cycle with only forward rates has steady-state probabilities
\begin{equation}
P_1^{\text{ss}} = \left[\left(\BAREONETWO\right)^{-1} + \left(\BARETWOTHREE\right)^{-1}e^{\DISS_{12} - \DISS_{23}} + \left(\BARETHREEONE\right)^{-1}e^{\DISS_{12} - \DISS_{31}}\right]^{-1} \ ,
\end{equation}
with cyclic permutation of states giving $P_2^{\text{ss}}$ and $P_3^{\text{ss}}$. The resulting steady-state flux is
\begin{equation}
J = \left[\left(\BAREONETWO e^{\DISS_{12}}\right)^{-1} + \left(\BARETWOTHREE e^{\DISS_{23}}\right)^{-1} + \left(\BARETHREEONE e^{\DISS_{31}}\right)^{-1}\right]^{-1} \ .
\end{equation}
Solving for $\partial J/\partial\DISS_{12} = \partial J/\partial\DISS_{23} = 0$ gives the optimal allocation
\begin{subequations}
\begin{align}
\label{eq:HighEntropyProduction_app}
\DOPT_{12} &= \frac{1}{3}\TOT + \frac{1}{3}\ln\frac{\BARETWOTHREE\BARETHREEONE}{\left(\BAREONETWO\right)^2} \ , \\
\DOPT_{23} &= \frac{1}{3}\TOT + \frac{1}{3}\ln\frac{\BARETHREEONE\BAREONETWO}{\left(\BARETWOTHREE\right)^2} \ , \\
\DOPT_{31} &= \frac{1}{3}\TOT + \frac{1}{3}\ln\frac{\BAREONETWO\BARETWOTHREE}{\left(\BARETHREEONE\right)^2} \ .
\end{align}
\end{subequations}

\subsection{Three-state flux for low $\DISS_{\text{tot}}$}
\label{subsec:ThreeStateLow_App}
Substituting \powerstrokenew rate constants into Eq.~\eqref{eq:ThreeStateFlux}
and solving for $\partial J/\partial\DISS_{12} = \partial J/\partial\DISS_{23} = 0$, subject to fixed $\TOT = \DISS_{12} + \DISS_{23} + \DISS_{31}$, gives
\begin{subequations}
\begin{align}
e^{2\DOPT_{12}} &= \BARETHREEONE\frac{(\BAREONETWO)^{-1} + (\BARETWOTHREE)^{-1} e^{-\DOPT_{23}}}{1 + e^{\DOPT_{23}}}e^{\TOT} \ , \\
e^{2\DOPT_{23}} &= \frac{1}{\BARETWOTHREE}\frac{1 + e^{-\DOPT_{12}}}{(\BAREONETWO)^{-1} + (\BARETHREEONE)^{-1} e^{\DOPT_{12}}}e^{\TOT} \ .
\end{align}
\end{subequations}
For small $\TOT$, $e^{\TOT} \simeq 1$, giving
\begin{subequations}
\begin{align}
\label{eq:LowDissipationOptApproxA}
e^{2\DOPT_{12}} &= \BARETHREEONE\frac{(\BAREONETWO)^{-1} + (\BARETWOTHREE)^{-1} e^{-\DOPT_{23}}}{1 + e^{\DOPT_{23}}} \ , \\
\label{eq:LowDissipationOptApproxB}
e^{2\DOPT_{23}} &= \frac{1}{\BARETWOTHREE}\frac{1 + e^{-\DOPT_{12}}}{(\BAREONETWO)^{-1} + (\BARETHREEONE)^{-1} e^{\DOPT_{12}}} \ .
\end{align}
\end{subequations}
Substituting Eq.~\eqref{eq:LowDissipationOptApproxB} into Eq.~\eqref{eq:LowDissipationOptApproxA} gives
\begin{equation}
\DOPT_{12} = \frac{1}{2}\ln\frac{\BARETHREEONE}{\BAREONETWO} \ .
\end{equation}
Similar derivations yield
\begin{subequations}
\begin{align}
\DOPT_{23} = \frac{1}{2}\ln\frac{\BAREONETWO}{\BARETWOTHREE} \ , \\
\DOPT_{31} = \frac{1}{2}\ln\frac{\BARETWOTHREE}{\BARETHREEONE} \ .
\end{align}
\end{subequations}

\section{\BrownianRatchetnew Scheme}
\label{sec:br_app}

\subsection{Two-state flux}
Given rate constants $\RCF$ and $\RCR$, the steady-state flux is Eq.~\eqref{eq:TwoStateFluxApp}. Substituting \brownianratchetnew rate constants $\RCF = \BARE$ and $\RCR = \BARE e^{-\DISS_{ij}}$ gives
\begin{equation}
\label{eq:FluxTwoStateBR}
J = \frac{1 - e^{-\TOT}}{(\BARETWOONE)^{-1}\left(1 + e^{-\DISS_{12}}\right) + (\BAREONETWO)^{-1}\left(1 + e^{-\DISS_{21}}\right)} \ .
\end{equation}
Solving $\MD J/\MD\DISS_{12}=0$ subject to fixed $\TOT = \DISS_{12} + \DISS_{21}$ gives
\begin{subequations}
\begin{align}
\DOPT_{12} &= \frac{1}{2}\TOT - \frac{1}{2}\ln\frac{\BARETWOONE}{\BAREONETWO} \\
\label{eq:FluxTwoStateBROpt}
J^* &= \frac{1 - e^{-\TOT}}{(\BAREONETWO)^{-1} + (\BARETWOONE)^{-1} + 2(\BAREONETWO\BARETWOONE e^{\TOT})^{-1/2}} \ .
\end{align}
\end{subequations}

Following similar steps as in Section~\ref{subsec:TwoState_App}, we find for small $|\delta| = |\DISS_{12} - \DOPT_{12}|$,
\begin{equation}
\label{eq:FluxCloseApproxBR}
\frac{J}{J^*} \simeq 1 - \left(\frac{\BAREONETWO + \BARETWOONE}{\sqrt{\BAREONETWO\BARETWOONE}}e^{\TOT/2} + 2\right)^{-1}\delta^2 \ ,
\end{equation}
and for large $|\delta| = |\DISS_{12} - \DOPT_{12}|$,
\begin{equation}
\label{eq:FluxFarApproxBR}
\frac{J}{J^*} \simeq \left(\frac{\BAREONETWO + \BARETWOONE}{\sqrt{\BAREONETWO\BARETWOONE}}e^{\TOT/2} + 2\right)e^{-|\delta|} \ .
\end{equation}
Fig.~\ref{fig:TwoStateSensitivity} compares Eqs.~\eqref{eq:FluxCloseApproxBR} and \eqref{eq:FluxFarApproxBR} to exact $J/J^*$, showing good agreement in the expected regimes.

For the two-state cycle, $\DOPTPS_{12} = \DOPTBR_{21}$ and $\DOPTPS_{21} = \DOPTBR_{12}$. This gives
\begin{subequations}
\begin{align}
\frac{J_{12+}^{*,\text{FL}}}{J_{21-}^{*,\text{FL}}} &= \frac{\BAREONETWO}{\BARETWOONE}e^{\DOPTPS_{12}} \ ,\\
\frac{J_{12+}^{*,\text{RL}}}{J_{21-}^{*,\text{RL}}} &= \frac{\BAREONETWO}{\BARETWOONE} e^{\DOPTBR_{21}}\ .
\end{align}
\end{subequations} 
Substituting $\DOPTPS_{12} = \DOPTBR_{21}$ gives
\begin{equation}
\label{eq:psbrratios}
\frac{J_{12+}^{*,\text{FL}}}{J_{21-}^{*,\text{FL}}} = \frac{J_{12+}^{*,\text{RL}}}{J_{21-}^{*,\text{RL}}} \ .
\end{equation}
Because $\DOPTPS_{12} = \frac{1}{2}\TOT + \frac{1}{2}\ln(\BARETWOONE/\BAREONETWO)$, the ratios in Eq.~\eqref{eq:psbrratios} are $\sqrt{\BAREONETWO/\BARETWOONE}e^{\TOT/2}$.

\subsection{Three-state flux for high $\TOT$}
\label{subsec:ThreeStateHighBR_App}
Substituting \brownianratchetnew rate constants into Eq.~\eqref{eq:ThreeStateFlux}
and solving for $\partial J/\partial\DISS_{12} = \partial J/\partial\DISS_{23} = 0$, subject to fixed $\TOT = \DISS_{12} + \DISS_{23} + \DISS_{31}$, gives
\begin{subequations}
\begin{align}
\label{eq:BRThreeStateHighA}
e^{-2\DOPT_{12}} &= \frac{1}{\BAREONETWO}\frac{1 + e^{\DOPT_{23}}}{(\BARETWOTHREE)^{-1} + (\BARETHREEONE)^{-1} e^{-\DOPT_{23}}}e^{-\TOT} \ , \\
\label{eq:BRThreeStateHighB}
e^{-2\DOPT_{23}} &= \BARETHREEONE\frac{(\BARETWOTHREE)^{-1} + (\BAREONETWO)^{-1} e^{\DOPT_{12}}}{1 + e^{-\DOPT_{12}}}e^{-\TOT} \ .
\end{align}
\end{subequations}
For high $\TOT$, these two equations are satisfied by
\begin{equation}
\DOPT_{12} = \frac{1}{3}\TOT + \frac{1}{3}\ln\frac{\BAREONETWO\BARETHREEONE}{\left(\BARETWOTHREE\right)^2} \ .
\end{equation}

\subsection{Three-state flux for low $\TOT$}
\label{subsec:ThreeStateLowBR_App}
Approximating $e^{-\TOT} \simeq 1$ in Eqs.~\eqref{eq:BRThreeStateHighA} and \eqref{eq:BRThreeStateHighB} gives
\begin{subequations}
\begin{align}
\label{eq:BRThreeStateLowA}
e^{-2\DOPT_{12}} &= \frac{1}{\BAREONETWO}\frac{1 + e^{\DOPT_{23}}}{(\BARETWOTHREE)^{-1} + (\BARETHREEONE)^{-1} e^{-\DOPT_{23}}} \ , \\
\label{eq:BRThreeStateLowB}
e^{-2\DOPT_{23}} &= \BARETHREEONE \frac{(\BARETWOTHREE)^{-1} + (\BAREONETWO)^{-1} e^{\DOPT_{12}}}{1 + e^{-\DOPT_{12}}} \ .
\end{align}
\end{subequations}
Substituting Eq.~\eqref{eq:BRThreeStateLowB} into Eq.~\eqref{eq:BRThreeStateLowA} gives
\begin{equation}
\DOPT_{12} = \frac{1}{2}\ln\frac{\BAREONETWO}{\BARETWOTHREE} \ .
\end{equation}

\section{Experimentally parameterized models}
\label{sec:comparison_app}

Table~\ref{tab:parameters} shows Hwang and Hyeon's~\cite{hwang17}
two-state parameterization of the forward and reverse rate constants for catalase, urease, alkaline phosphatase (AP), triose phosphate isomerase (TPI), and kinesin. The dissipation allocation $\DISS_{12}$ and $\DISS_{21}$ from these rate constants is compared to the optimal dissipation allocations predicted for \powerstrokenew cycles, $\DISS_{ij}^{*,\text{FL}}$, and \brownianratchetnew cycles, $\DISS_{ij}^{*,\text{RL}}$.
Fig.~6 summarizes the comparison of experimental fit, forward labile prediction, and even allocation.

\begin{table}[tbp]
  \centering
  \begin{tabular}{c||ccccc}
  & Catalase & Urease & AP & TPI & Kinesin\\
  \hline
  \hline
  \rule{0pt}{3ex}$\RCFONETWO$ & $5.8\times10^4$ & $1.7\times10^4$ & $1.5\times10^5$ & $1.7\times10^5$ & $2.2\times10^{3}$\\
  $\RCRONETWO$ & $2.2\times10^{-13}$ & $7.4\times10^{-7}$ & $4\times10^{-4}$ & $4.2\times10^3$ & $5.5\times10^{-1}$\\
  $\RCFTWOONE$ & $6.2\times10^6$ & $3\times10^5$ & $1.6\times10^5$ & $1.8\times10^5$ & $9.9\times10^1$\\
  $\RCRTWOONE$ & $6.1\times10^6$ & $2.8\times10^5$ & $1.4\times10^4$ & $1.3\times10^4$ & $9.2\times10^{-2}$\\
  \hline
  \rule{0pt}{3ex}$\DISS_{12}$ & $40$ & $24$ & $20$ & $3.7$ & $8.3$\\
  $\DISS_{12}^{\text{even}}$ & $20$ & $12$ & $10$ & $3.2$ & $7.7$\\
  $\DISS_{12}^{*,\text{FL}}$ & $42.4$ & $25.3$ & $18.6$ & $3.7$ & $6.6$\\
  $\DISS_{12}^{*,\text{RL}}$ & $17.7$ & $6.6$ & $7.5$ & $2.6$ & $9.1$\\
  \hline
  \rule{0pt}{3ex}$\DISS_{21}$ & $0.02$ & $0.07$ & $0.13$ & $2.6$ & $7$\\
  $\DISS_{21}^{\text{even}}$ & $20$ & $12$ & $10$ & $3.2$ & $7.7$ \\
  $\DISS_{21}^{*,\text{FL}}$ & $-2.4$ & $-1.3$ & $-1.2$ & $2.6$ & $8.4$\\
  $\DISS_{21}^{*,\text{RL}}$ & $22.3$ & $13.4$ & $9.9$ & $3.7$ & $5.9$ \\
  \end{tabular}
  \caption{\label{tab:parameters} 
  {\bf Comparing theoretical predictions and experimental fits of dissipation allocation in two-state enzymatic models}.
  $\RCFONETWO$, $\RCRONETWO$, $\RCFTWOONE$, $\RCRTWOONE$ are from Table 1 of Hwang and Hyeon~\cite{hwang17}. $\DISS_{12}$ and $\DISS_{21}$ are calculated using Eq.~\eqref{eq:ratio}, $\DISS_{ij}^{*,\text{FL}}$ using Eq.~\eqref{eq:TwoStateFluxOpt}, and $\DISS_{ij}^{*,\text{RL}}$ using Eq.~\eqref{eq:TwoStateOpt}. Without loss of generality, we adopt the convention that transition 12 is the one with higher dissipation.
AP, alkaline phosphatase; TPI, triose phosphate isomerase.} 
\end{table}

For catalase, urease, and AP, the \powerstrokenew prediction is quite close to the parameters from \cite{hwang17}, while the \brownianratchetnew prediction is qualitatively different. For TPI, the \powerstrokenew prediction is a very close match to the parameters from \cite{hwang17}, but \brownianratchetnew prediction is not qualitatively different. For kinesin, neither the \powerstrokenew nor \brownianratchetnew predictions are clearly a better match for the parameters of \cite{hwang17}. 

\begin{table}[tbp]
  \centering
  \begin{tabular}{c||ccc}
  & 12 & 23 & 31 \\
  \hline
  \hline
  \rule{0pt}{3ex}$\RCF$ & 3000 & 570 & 57\\
  $\RCR$ & 68 & 0.2 & 0.02\\
  \hline
  \rule{0pt}{3ex}$\DISS_{ij}$ & 4 & 8 & 8\\
  $\DISS_{ij}^{\text{even}}$ & 6.7 & 6.7 & 6.7\\
  $\DISS_{ij}^{*,\text{FL}}$ & 2 & 7.8 & 10.1\\
  $\DISS_{ij}^{*,\text{RL}}$ & 6.5 & 8.8 & 4.8\\
  \end{tabular}
  \caption{\label{tab:kinesinparametersthree}
  {\bf Comparing theoretical predictions and experimental fits of dissipation allocation in a three-state kinesin model}.
  $\RCRONETWO$, $\RCFTWOTHREE$, and $\RCFTHREEONE$ are directly from Clancy, \emph{et al}~\cite{clancy11} for a three-state main cycle model of mutant kinesin. $\RCFONETWO$, $\RCRTWOTHREE$, $\RCRTHREEONE$, and $\DISS_{ij}$ calculated as described in text. $\DISS_{ij}^{*,\text{FL}}$ predicted from Eq.~\eqref{eq:HighEntropyProduction}, $\DISS_{ij}^{*,\text{RL}}$ from Eq.~\eqref{eq:BR_A}.
}
\end{table}

Table~\ref{tab:kinesinparametersthree} shows the three-state parameterization of kinesin from Clancy \emph{et al}~\cite{clancy11}. Typical physiological ATP concentrations in the low millimolars~\cite{huang10} motivate the approximation [ATP]$\sim$1mM, giving $\RCFONETWO$, and hence $\DISS_{12} \simeq 4$. The second and third transitions are considered `irreversible,' so we assume that the remaining dissipation budget (from the 20 $k_{\text{B}}T$ free energy provided by ATP hydrolysis) is evenly split to these two transitions, so that $\DISS_{23} = \DISS_{31} = 8$, providing values for $\RCRTWOTHREE$ and $\RCRTHREEONE$. The \powerstrokenew prediction is closer to the parameters from~\cite{clancy11} than the \brownianratchetnew prediction.

\begin{table}[tbp]
  \centering
  \begin{tabular}{c||cccc}
  & 12 & 23 & 34 & 41\\
  \hline
  \hline
  \rule{0pt}{3ex}$\RCF$ & 3000 & 600 & 400 & 190\\
  $\RCR$ & 20 & 1.4 & 1.7 & 120\\
  \hline
  \rule{0pt}{3ex}$\DISS_{ij}$ & 5 & 6.1 & 5.5 & 1.6\\
  $\DISS_{ij}^{\text{even}}$ & 4.6 & 4.6 & 4.6 & 4.6\\
  $\DISS_{ij}^{*,\text{FL}}$ & Third & First & Second & Fourth\\
  $\DISS_{ij}^{*,\text{RL}}$ & First & Second & Third & Fourth\\
  \end{tabular}
  \caption{\label{tab:kinesinparametersfour} 
  {\bf Comparing theoretical predictions and experimental fits of dissipation allocation in a four-state kinesin model}.
  All constants except $\RCFONETWO$ are directly from Hwang and Hyeon's parameterization of a four-state model for kinesin~\cite{hwang17}. $\RCFONETWO$ calculation and $\DOPT_{ij}$ ranking described in text.}
\end{table}

Table~\ref{tab:kinesinparametersfour} shows the four-state parameterization of kinesin from Hwang and Hyeon~\cite{hwang17}. $\RCFONETWO$ assumes an ATP concentration of 1mM. Since we do not have quantitative predictions for a four-state cycle, we rank the optimal order for dissipation assigned for a \powerstrokenew scheme with more dissipation allocated to smaller reverse rate constants (`First' indicates largest dissipation), and for a \brownianratchetnew scheme rank optimal dissipation order by assigning more dissipation to larger forward rate constants. 
\Powerstrokenew 
predicts the correct ordering of transition dissipations, whereas the \brownianratchetnew does not.

	\bibliography{BrownSivak_OptimizingDissipation}

\end{document}